\documentclass[12pt]{article}
\usepackage{authblk}
\usepackage[dvips]{graphicx}
\graphicspath{{noiseimages/}}
\usepackage{xcolor}
\usepackage{tikz}
\usepackage{pgfplots}
\usetikzlibrary{intersections, pgfplots.fillbetween}
\usetikzlibrary{decorations.pathmorphing}
\usetikzlibrary{snakes}
\usepackage{amsfonts}
\usepackage[english]{babel}
\usepackage{color}
\usepackage{amsmath,amssymb}

\def\hybrid{\topmargin 0pt      \oddsidemargin 0pt
        \headheight 0pt \headsep 0pt
       \voffset-1cm
        \textwidth 6.25in       
       \textheight 9.5in       
        \marginparwidth 0.0in
        \parskip 5pt plus 1pt   \jot = 1.5ex}
\catcode`\@=11
\def\marginnote#1{}

\newcount\hour
\newcount\minute
\newtoks\amorpm
\hour=\time\divide\hour by60
\minute=\time{\multiply\hour by60 \global\advance\minute by-\hour}
\edef\standardtime{{\ifnum\hour<12 \global\amorpm={am}%
        \else\global\amorpm={pm}\advance\hour by-12 \fi
        \ifnum\hour=0 \hour=12 \fi
        \number\hour:\ifnum\minute<10 0\fi\number\minute\the\amorpm}}
\edef\militarytime{\number\hour:\ifnum\minute<10 0\fi\number\minute}

\def\draftlabel#1{{\@bsphack\if@filesw {\let\thepage\relax
   \xdef\@gtempa{\write\@auxout{\string
      \newlabel{#1}{{\@currentlabel}{\thepage}}}}}\@gtempa
   \if@nobreak \ifvmode\nobreak\fi\fi\fi\@esphack}
        \gdef\@eqnlabel{#1}}
\def\@eqnlabel{}
\def\@vacuum{}
\def\draftmarginnote#1{\marginpar{\raggedright\scriptsize\tt#1}}

\def\draftlabel#1{{\@bsphack\if@filesw {\let\thepage\relax
   \xdef\@gtempa{\write\@auxout{\string
      \newlabel{#1}{{\@currentlabel}{\thepage}}}}}\@gtempa
   \if@nobreak \ifvmode\nobreak\fi\fi\fi\@esphack}
        \gdef\@eqnlabel{#1}}
\def\@eqnlabel{}
\def\@vacuum{}
\def\draftmarginnote#1{\marginpar{\raggedright\scriptsize\tt#1}}

\def\draft{\oddsidemargin -.5truein
        \def\@oddfoot{\sl preliminary draft \hfil
        \rm\thepage\hfil\sl\today\quad\militarytime}
        \let\@evenfoot\@oddfoot \overfullrule 3pt
        \let\label=\draftlabel
        \let\marginnote=\draftmarginnote
   \def\@eqnnum{(\theequation)\rlap{\kern\marginparsep\tt\@eqnlabel}%
\global\let\@eqnlabel\@vacuum}  }

\def\numberbysection{\@addtoreset{equation}{section}
        \def\theequation{\thesection.\arabic{equation}}}

\def\underline#1{\relax\ifmmode\@@underline#1\else
        $\@@underline{\hbox{#1}}$\relax\fi}

\def\titlepage{\@restonecolfalse\if@twocolumn\@restonecoltrue\onecolumn
     \else \newpage \fi \thispagestyle{empty}\c@page\z@
        \def\thefootnote{\fnsymbol{footnote}} }

\def\endtitlepage{\if@restonecol\twocolumn \else  \fi
        \def\thefootnote{\arabic{footnote}}
        \setcounter{footnote}{0}}  
\relax

\numberbysection
\hybrid

\newfont{\Bbbb}{msbm7 scaled 1\@ptsize00}
\newcommand{\Ann}{\mathbb A}
\newcommand{\CC}{\mathbb C}
\newcommand{\SSS}{\mathbb{S}}

\newcommand{\DDD}{\raise-1pt\hbox{$\mbox{\Bbbb D}$}}
\newcommand{\HH}{\mathbb{H}}

\newcommand{\RR}{\mathbb{R}}

\newcommand{\UU}{\mathbb{U}}
\newcommand{\UUU}{\raise-1pt\hbox{$\mbox{\Bbbb U}$}}

\newcommand{\z}{\raise-1pt\hbox{$\mbox{\Bbbb Z}$}}

\def\beq{\begin{equation}}
\def\eeq{\end{equation}}
\def\p{\partial}

\newtheorem{lemma-definition}{Lemma-Definition}[section]

\newcommand{\calK}{\mathcal{K}}
\newcommand{\Comp}{\mathbb{C}}
\newcommand{\der}{\partial}
\renewcommand{\Im}{\operatorname{Im}}
\renewcommand{\Re}{\operatorname{Re}}
\newcommand{\Real}{\mathbb{R}}
\renewcommand{\setminus}{\smallsetminus}
\newcommand{\sfR}{\mathsf{R}}

\begin{document}

\begin{titlepage}



\title{L\"owner equations and\\ reductions of dispersionless hierarchies}

\author{V.~Akhmedova\thanks{Kharkevich Institute for Information Transmission Problems, Bolshoy Karetny per. 19, build.1, Moscow, 127051, Russia,
e-mail: valeria-58@yandex.ru}
\and T.~Takebe 
\thanks{
Faculty of Mathematics, National Research University Higher School of
Ecnomics, Russian Federation, Usacheva str., 6, Moscow, 119048;
HSE-Skoltech International Laboratory of Representation Theory and
Mathematical Physics, Usacheva str., 6, Moscow, 119048, Russia, e-mail: ttakebe@hse.ru}
\and A.~Zabrodin
\thanks{
Skolkovo Institute of Science and Technology, 143026, Moscow, Russia
and
Institute of Biochemical Physics, Kosygina str. 4, 119334, Moscow, Russia
and
ITEP NRC KI, 25
B.Cheremushkinskaya, Moscow 117218, Russia,
e-mail: zabrodin@itep.ru}}

\date{October 2020}
\maketitle

\vspace{-9cm} \centerline{ \hfill ITEP-TH-19/20}\vspace{9cm}

\vspace{-9cm} \centerline{ \hfill IITP-TH-14/20}\vspace{9cm}

\begin{abstract}

The equations of L\"owner type can be derived in two very different contexts:
one of them is complex analysis and the theory of parametric conformal maps and the
other one is the theory of integrable systems. In this paper we compare the both approaches.
After recalling the derivation of L\"owner equations based on complex analysis
we review one- and multi-variable reductions of dispersionless integrable hierarhies 
(dKP, dBKP, dToda, and dDKP). The one-vaiable reductions 
are described by solutions of different versions of L\"owner
equation: chordal (rational) for dKP, quadrant for dBKP,
radial (trigonometric) for dToda and 
elliptic for DKP. We also discuss multi-variable reductions which are
given by a system of 
L\"owner equations supplemented by a system of partial differential 
equations of hydrodynamic type. The solvability of the hydrodynamic type system 
can be proved by means of the 
generalized hodograph method.

\end{abstract}

\end{titlepage}

\vspace{5mm}

%

\tableofcontents

\vspace{5mm} 

\section{Introduction}

Around the turn of the millennium it turned out that the hierarchies of integrable
partial differential equations are related to the theory of univalent functions 
(i.e., to the Riemann mapping theorem) unexpectedly deeply. To wit, dispersionless 
limits of the Kadomtsev-Petviashvili (dKP) and the 2D Toda (dToda) hierarchies 
were shown to be closely related to conformal maps of domains in the complex plane. 

This relation was developed in two seemingly different directions. One of them treats 
equations of the dToda hierarchy as governing equations for conformal maps of plane
domains with smooth boundary as functions of their harmonic moments \cite{MWWZ00,WZ00}.
Another one is related to conformal maps of domains with slits (slit domains). In the
seminal papers \cite{GT1,GT2} reductions of the dKP hierarchy were studied and it was discovered
that they are classified by solutions of a L\"owner-type differential equation which 
characterizes one-parameter families of conformal mappings of domains with a growing slit
and a fixed reference domain \cite{L1923}. 
Later this important observation was extended to hierarchies 
of other types and other types of L\"owner-like equations 
\cite{Manas1,Manas2,TT06,TTZ06,T13,akh-zab:14-1}. 

The aim of this paper is to review reductions of the 
dispersionless hierarhies such as dKP, dToda as well as of the B- and D- versions of the 
dKP hierarchy (dBKP and dDKP) within a unified framework of Hirota's approach and to elucidate
their deep relation with different types of the L\"owner equation (respectively,
chordal, radial, quadrant and elliptic). A conceptual 
understanding of mathematical origin of this relation is still missing. 

The meaning of reduction is as follows. 
Infinite hierarchies of partial differential equations contain an infinite number of
independent variables (``times'') and an infinite number of dependent variables. 
The simplest possible reduction is a reduction to just one dependent variable which
depends on all the times, and all other dependent
variables become functions of it (a one-variable reduction). 
It appears that the one-variable reductions are described
by solutions of a single L\"owner equation with a driving 
function which characterizes the reduction. 
Geometrically, the driving function 
characterizes the shape of the slit and 
the single dependent variable is a parameter along the slit. 
One can also consider multi-variable 
($N$-variable) reductions
when there are $N$ dependent variables. In this case the reduction is described
by a system of $N$ L\"owner equations with $N$ driving functions 
and certain compatibility conditions (the Gibbons-Tsarev 
equations \cite{GT1}) appear. 

In section 2 we review the different types of 
L\"owner equations from the point of view of complex analysis and conformal mappings.
We would like to stress that the hierarchical sequence of L\"owner equations
$$
\mbox{{\it chordal $\longrightarrow$ radial $\longrightarrow$ elliptic}}
$$
corresponds to the types of functions
$$
\mbox{{\it rational $\longrightarrow$ trigonometric $\longrightarrow$ elliptic}}
$$
and, on the side of dispersionless integrable hierarchies, to
$$
\mbox{{\it dKP $\longrightarrow$ dToda $\longrightarrow$ dDKP}.}
$$

In section \ref{section:one}
we consider dispersionless integrable hierarchies: dKP, dBKP, dToda and dDKP
(the order corresponds to growing complexity).
In each case we start from the generating equation for the dispersionless
limit of the tau-function (coming from the Hirota bilinear equations) and 
in the cases of dKP, dBKP and dToda prove
that it is equivalent to the more familiar Lax formulation. (For the dDKP hierarchy
in the elliptic form the Lax formulation is not known.) For each case, 
the one-variable reduction is considered and the corresponding L\"owner equation
(chordal, quadrant, radial and elliptic) is derived as the consistency condition 
of the reduction with the structure of the infinite hierarchy.  

Section \ref{section:multi} is devoted to multi-variable reductions characterized by
a system of L\"owner equations. We derive the compatibility conditions for them which are
the Gibbons-Tsarev equations. In the case of reduction, the infinite hierarchy is reduced 
to a finite system of differential equations of hydrodynamic type for a finite number of
dependent variables. This system is implicitly 
solved by means of the generalized hodograph method developed by Tsarev \cite{tsa:90}.

\section{L\"owner equations}

The conformal map of a domain with a slit of arbitrary shape to a reference domain
(say, the upper half plane or the unit circle) satisfy certain differential equation 
as a function of a parameter characterizing the slit. Such differential equations are called
equations of the L\"owner type. 

\subsection{Chordal L\"owner equation}

\label{section:chordal}

The simplest equation of this type, now called chordal L\"owner
equation, was first obtained in \cite{KSS68}. 
Let $\Gamma$ be a smooth curve $\Gamma:[0,+\infty) \to \HH$ in the upper
half plane $\HH$ starting from a point on the real axis,
$\Gamma(0)\in\Real$, and $\Gamma_t$ be its arc,
$\Gamma_t:=\Gamma([0,t])$.
We assume that, if the curve touches the real axis, it goes off and enters
$\HH$ immediately.

According to the Riemann mapping theorem, there exists a unique
univalent conformal map $g(z, t)$ from $\HH \setminus \Gamma_t$ to $\HH$
normalized by the condition
\begin{equation*}
g(z,t)=z +u(t)z^{-1} + O(z^{-2}), \quad z\to \infty , \quad t>0. 
\end{equation*}
The parameter $t$ along the curve is sometimes referred to as ``time''
and the coefficient $u(t)$ as the ``capacity'' of the domain $\HH
\setminus \Gamma_t$.

It can be shown (see, for example, \cite{kag-nie:04}) that $u(t)$ is a
continuous increasing function of $t$, which makes it possible to assume
$u(t)=t$ by reparametrization. In this parametrization, $g(z,t)$ has the
following form:
\begin{equation}
g(z,t)=z +tz^{-1} + O(z^{-2}), \quad z\to \infty , \quad t>0. 
\label{g(z,t)} 
\end{equation}
%
%

\begin{figure}[h!]
\begin{tikzpicture}[scale=1]

\draw ( -8.0 ,-1/5) to ( -7.8 ,0);
\draw ( -7.8 ,-1/5) to ( -7.6 ,0);
\draw ( -7.6 ,-1/5) to ( -7.3999999999999995 ,0);
\draw ( -7.4 ,-1/5) to ( -7.2 ,0);
\draw ( -7.2 ,-1/5) to ( -7.0 ,0);
\draw ( -7.0 ,-1/5) to ( -6.8 ,0);
\draw ( -6.8 ,-1/5) to ( -6.6 ,0);
\draw ( -6.6 ,-1/5) to ( -6.3999999999999995 ,0);
\draw ( -6.4 ,-1/5) to ( -6.2 ,0);
\draw ( -6.2 ,-1/5) to ( -6.0 ,0);
\draw ( -6.0 ,-1/5) to ( -5.8 ,0);
\draw ( -5.8 ,-1/5) to ( -5.6 ,0);
\draw ( -5.6 ,-1/5) to ( -5.3999999999999995 ,0);
\draw ( -5.4 ,-1/5) to ( -5.2 ,0);
\draw ( -5.2 ,-1/5) to ( -5.0 ,0);
\draw ( -5.0 ,-1/5) to ( -4.8 ,0);
\draw ( -4.8 ,-1/5) to ( -4.6 ,0);
\draw ( -4.6 ,-1/5) to ( -4.3999999999999995 ,0);
\draw ( -4.4 ,-1/5) to ( -4.2 ,0);
\draw ( -4.2 ,-1/5) to ( -4.0 ,0);
\draw ( -4.0 ,-1/5) to ( -3.8 ,0);
\draw ( -3.8 ,-1/5) to ( -3.5999999999999996 ,0);
\draw ( -3.6 ,-1/5) to ( -3.4 ,0);
\draw ( -3.4 ,-1/5) to ( -3.1999999999999997 ,0);
\draw ( -3.2 ,-1/5) to ( -3.0 ,0);
\draw ( -3.0 ,-1/5) to ( -2.8 ,0);
\draw ( -2.8 ,-1/5) to ( -2.5999999999999996 ,0);
\draw ( -2.6 ,-1/5) to ( -2.4 ,0);
\draw ( -2.4 ,-1/5) to ( -2.1999999999999997 ,0);
\draw ( -2.2 ,-1/5) to ( -2.0 ,0);


\draw ( 1.6 ,-1/5) to ( 1.8 ,0);
\draw ( 1.8 ,-1/5) to ( 2.0 ,0);
\draw ( 2.0 ,-1/5) to ( 2.2 ,0);
\draw ( 2.2 ,-1/5) to ( 2.4000000000000004 ,0);
\draw ( 2.4 ,-1/5) to ( 2.6 ,0);
\draw ( 2.6 ,-1/5) to ( 2.8000000000000003 ,0);
\draw ( 2.8 ,-1/5) to ( 3.0 ,0);
\draw ( 3.0 ,-1/5) to ( 3.2 ,0);
\draw ( 3.2 ,-1/5) to ( 3.4000000000000004 ,0);
\draw ( 3.4 ,-1/5) to ( 3.6 ,0);
\draw ( 3.6 ,-1/5) to ( 3.8000000000000003 ,0);
\draw ( 3.8 ,-1/5) to ( 4.0 ,0);
\draw ( 4.0 ,-1/5) to ( 4.2 ,0);
\draw ( 4.2 ,-1/5) to ( 4.4 ,0);
\draw ( 4.4 ,-1/5) to ( 4.6000000000000005 ,0);
\draw ( 4.6 ,-1/5) to ( 4.8 ,0);
\draw ( 4.8 ,-1/5) to ( 5.0 ,0);
\draw ( 5.0 ,-1/5) to ( 5.2 ,0);
\draw ( 5.2 ,-1/5) to ( 5.4 ,0);
\draw ( 5.4 ,-1/5) to ( 5.6000000000000005 ,0);
\draw ( 5.6 ,-1/5) to ( 5.8 ,0);
\draw ( 5.8 ,-1/5) to ( 6.0 ,0);
\draw ( 6.0 ,-1/5) to ( 6.2 ,0);
\draw ( 6.2 ,-1/5) to ( 6.4 ,0);
\draw ( 6.4 ,-1/5) to ( 6.6000000000000005 ,0);
\draw ( 6.6 ,-1/5) to ( 6.8 ,0);
\draw ( 6.8 ,-1/5) to ( 7.0 ,0);
\draw ( 7.0 ,-1/5) to ( 7.2 ,0);
\draw ( 7.2 ,-1/5) to ( 7.4 ,0);
\draw (-8,0) to (-2,0);
\draw (1.6,0) to (7.4,0);

\node at (-8,2.6)  {\LARGE $\mathbb {H}$ \normalsize}; 
\node at (2,2.6)  {\LARGE $\mathbb {H}$ \normalsize}; 
\draw [thick, ->] (-1.05,1) -- node[midway, above] {\Large $g(z,t)$ \normalsize} (0.25,1);

\draw (-5,0) node[circle,fill,inner sep=1.5pt]{} to[out=90,in=-90] (-4,1) to[out=90,in=-90] (-4.5,2) to[out=90,in=-135] (-4,2.5) node[circle,fill,inner sep=1.5pt,label=right:$z_*$]{}; 
\node at (-3.5,1)  {\Large $\Gamma_t$ \normalsize};

\draw (4,0) node[circle,fill,inner sep=1.5pt]{};
\node at (4,-0.5) {\Large {$g(z_*,t)=\xi (t)$} \normalsize};
\end{tikzpicture}
\caption{}
    \label{fig1}
\end{figure}

It appears that there exists a continuous real-valued function $\xi(t)$ 
(called the driving function) such that 
$g(z,t)$ satisfies the differential equation
\beq\label{le0}
\frac{\p g(z, t)}{\p t}=\frac{1}{g(z, t)-\xi(t)}, \qquad g(z, 0)=z.
\eeq
This is the chordal L\"owner equation. The point $\xi (t)\in \RR$ 
is the image of the tip of the curve $\Gamma_t$. This is how the chordal 
L\"owner equation appears in complex analysis. It was first proved in 
\cite{KSS68} and rediscovered in the context of integrable systems
by Gibbons and Tsarev in \cite{GT2}. It became well-known when Schramm
\cite{S00} discovered it independently and studied random curves in the upper
half plane (in the celebrated SLE the driving function $\xi(t)$ is a Brownian motion).

\bigskip
%

\begin{figure}[h!]
\begin{tikzpicture}[scale=1]

\draw ( -8.0 ,-1/5) to ( -7.8 ,0);
\draw ( -7.8 ,-1/5) to ( -7.6 ,0);
\draw ( -7.6 ,-1/5) to ( -7.3999999999999995 ,0);
\draw ( -7.4 ,-1/5) to ( -7.2 ,0);
\draw ( -7.2 ,-1/5) to ( -7.0 ,0);
\draw ( -7.0 ,-1/5) to ( -6.8 ,0);
\draw ( -6.8 ,-1/5) to ( -6.6 ,0);
\draw ( -6.6 ,-1/5) to ( -6.3999999999999995 ,0);
\draw ( -6.4 ,-1/5) to ( -6.2 ,0);
\draw ( -6.2 ,-1/5) to ( -6.0 ,0);
\draw ( -6.0 ,-1/5) to ( -5.8 ,0);
\draw ( -5.8 ,-1/5) to ( -5.6 ,0);
\draw ( -5.6 ,-1/5) to ( -5.3999999999999995 ,0);
\draw ( -5.4 ,-1/5) to ( -5.2 ,0);
\draw ( -5.2 ,-1/5) to ( -5.0 ,0);
\draw ( -5.0 ,-1/5) to ( -4.8 ,0);
\draw ( -4.8 ,-1/5) to ( -4.6 ,0);
\draw ( -4.6 ,-1/5) to ( -4.3999999999999995 ,0);
\draw ( -4.4 ,-1/5) to ( -4.2 ,0);
\draw ( -4.2 ,-1/5) to ( -4.0 ,0);
\draw ( -4.0 ,-1/5) to ( -3.8 ,0);
\draw ( -3.8 ,-1/5) to ( -3.5999999999999996 ,0);
\draw ( -3.6 ,-1/5) to ( -3.4 ,0);
\draw ( -3.4 ,-1/5) to ( -3.1999999999999997 ,0);
\draw ( -3.2 ,-1/5) to ( -3.0 ,0);
\draw ( -3.0 ,-1/5) to ( -2.8 ,0);
\draw ( -2.8 ,-1/5) to ( -2.5999999999999996 ,0);
\draw ( -2.6 ,-1/5) to ( -2.4 ,0);
\draw ( -2.4 ,-1/5) to ( -2.1999999999999997 ,0);
\draw ( -2.2 ,-1/5) to ( -2.0 ,0);


\draw ( 1.6 ,-1/5) to ( 1.8 ,0);
\draw ( 1.8 ,-1/5) to ( 2.0 ,0);
\draw ( 2.0 ,-1/5) to ( 2.2 ,0);
\draw ( 2.2 ,-1/5) to ( 2.4000000000000004 ,0);
\draw ( 2.4 ,-1/5) to ( 2.6 ,0);
\draw ( 2.6 ,-1/5) to ( 2.8000000000000003 ,0);
\draw ( 2.8 ,-1/5) to ( 3.0 ,0);
\draw ( 3.0 ,-1/5) to ( 3.2 ,0);
\draw ( 3.2 ,-1/5) to ( 3.4000000000000004 ,0);
\draw ( 3.4 ,-1/5) to ( 3.6 ,0);
\draw ( 3.6 ,-1/5) to ( 3.8000000000000003 ,0);
\draw ( 3.8 ,-1/5) to ( 4.0 ,0);
\draw ( 4.0 ,-1/5) to ( 4.2 ,0);
\draw ( 4.2 ,-1/5) to ( 4.4 ,0);
\draw ( 4.4 ,-1/5) to ( 4.6000000000000005 ,0);
\draw ( 4.6 ,-1/5) to ( 4.8 ,0);
\draw ( 4.8 ,-1/5) to ( 5.0 ,0);
\draw ( 5.0 ,-1/5) to ( 5.2 ,0);
\draw ( 5.2 ,-1/5) to ( 5.4 ,0);
\draw ( 5.4 ,-1/5) to ( 5.6000000000000005 ,0);
\draw ( 5.6 ,-1/5) to ( 5.8 ,0);
\draw ( 5.8 ,-1/5) to ( 6.0 ,0);
\draw ( 6.0 ,-1/5) to ( 6.2 ,0);
\draw ( 6.2 ,-1/5) to ( 6.4 ,0);
\draw ( 6.4 ,-1/5) to ( 6.6000000000000005 ,0);
\draw ( 6.6 ,-1/5) to ( 6.8 ,0);
\draw ( 6.8 ,-1/5) to ( 7.0 ,0);
\draw ( 7.0 ,-1/5) to ( 7.2 ,0);
\draw ( 7.2 ,-1/5) to ( 7.4 ,0);
\draw (-8,0) to (-2,0);
\draw (1.6,0) to (7.4,0);

\node at (-8,2)  {\LARGE $\mathbb {H}$ \normalsize}; 
\node at (2,2)  {\LARGE $\mathbb {H}$ \normalsize}; 

\draw [thick, ->] (-1.05,1) -- node[midway, above] {\Large $g(z,t)$ \normalsize} (0.25,1);

\node at (-5,-0.5) {\large {$\xi$} \normalsize};
\draw (-5,0) node[circle,fill,inner sep=1.5pt]{} -- (-5,2) node[circle,fill,inner sep=1.5pt,label=right:$\xi+i\sqrt{2t}$]{}; 

\node at (4,-0.5) {\large {$\xi$} \normalsize};
\draw [very thick] (3,0) node[circle,fill,inner sep=1.5pt]{} -- (4,0) node[circle,fill,inner sep=1.5pt]{} -- (5,0) node[circle,fill,inner sep=1.5pt]{};
\end{tikzpicture}
\caption{}
    \label{fig2}
\end{figure}

For example, 
if $\Gamma_t = [\xi , \xi +i\sqrt{2t}]$ is the straight segment from $\xi \in \RR$ to
the point $\xi+i\sqrt{2t}$, then $\xi(t)=\xi =\mbox{const}$ and
\beq\label{le1}
g(z,t)=\xi + \sqrt{(z-\xi )^2 +2t}.
\eeq
Note that if $t\to 0$ and $|z-\xi |$ is bounded from below, then
\beq\label{le2}
g(z, t)=z+\frac{t}{z-\xi} +O(t^2).
\eeq

The origin of the equation (\ref{le0})
in a good situation (when, for example, the curve is smooth)
is explained as follows \cite{C05}. 
(This explanation is by no means rigorous but simple and
geometrically clear. For rigorous proofs we refer to, for example,
\cite{kag-nie:04}, \cite{d-mo-gum:13}.) 

\bigskip
Let the curve evolve for a ``time'' $t$ and then for a further short 
``time'' $s$, $s\to 0$. The image of $\HH \setminus 
\Gamma_{t+s}$ under $g(z,t+s)$ is $\HH$ while
the image of $\HH \setminus 
\Gamma_{t+s}$ under $g(z,t)$ is $\HH$ with a cut which is 
a short vertical segment starting from the point
$\xi (t)$ on the real 
axis:
The tip $z_*:=\Gamma(t)$ of the curve $\Gamma_{t}$ is mapped to
$g(z_*,t)=\xi(t)$ by the map $g(z,t)$ (Figure \ref{fig2}). 

\begin{figure}[h!]
\begin{tikzpicture}[scale=1]

\draw ( -8.0 ,-1/5) to ( -7.8 ,0);
\draw ( -7.8 ,-1/5) to ( -7.6 ,0);
\draw ( -7.6 ,-1/5) to ( -7.3999999999999995 ,0);
\draw ( -7.4 ,-1/5) to ( -7.2 ,0);
\draw ( -7.2 ,-1/5) to ( -7.0 ,0);
\draw ( -7.0 ,-1/5) to ( -6.8 ,0);
\draw ( -6.8 ,-1/5) to ( -6.6 ,0);
\draw ( -6.6 ,-1/5) to ( -6.3999999999999995 ,0);
\draw ( -6.4 ,-1/5) to ( -6.2 ,0);
\draw ( -6.2 ,-1/5) to ( -6.0 ,0);
\draw ( -6.0 ,-1/5) to ( -5.8 ,0);
\draw ( -5.8 ,-1/5) to ( -5.6 ,0);
\draw ( -5.6 ,-1/5) to ( -5.3999999999999995 ,0);
\draw ( -5.4 ,-1/5) to ( -5.2 ,0);
\draw ( -5.2 ,-1/5) to ( -5.0 ,0);
\draw ( -5.0 ,-1/5) to ( -4.8 ,0);
\draw ( -4.8 ,-1/5) to ( -4.6 ,0);
\draw ( -4.6 ,-1/5) to ( -4.3999999999999995 ,0);
\draw ( -4.4 ,-1/5) to ( -4.2 ,0);
\draw ( -4.2 ,-1/5) to ( -4.0 ,0);
\draw ( -4.0 ,-1/5) to ( -3.8 ,0);
\draw ( -3.8 ,-1/5) to ( -3.5999999999999996 ,0);
\draw ( -3.6 ,-1/5) to ( -3.4 ,0);
\draw ( -3.4 ,-1/5) to ( -3.1999999999999997 ,0);
\draw ( -3.2 ,-1/5) to ( -3.0 ,0);
\draw ( -3.0 ,-1/5) to ( -2.8 ,0);
\draw ( -2.8 ,-1/5) to ( -2.5999999999999996 ,0);
\draw ( -2.6 ,-1/5) to ( -2.4 ,0);
\draw ( -2.4 ,-1/5) to ( -2.1999999999999997 ,0);
\draw ( -2.2 ,-1/5) to ( -2.0 ,0);


\draw ( 1.6 ,-1/5) to ( 1.8 ,0);
\draw ( 1.8 ,-1/5) to ( 2.0 ,0);
\draw ( 2.0 ,-1/5) to ( 2.2 ,0);
\draw ( 2.2 ,-1/5) to ( 2.4000000000000004 ,0);
\draw ( 2.4 ,-1/5) to ( 2.6 ,0);
\draw ( 2.6 ,-1/5) to ( 2.8000000000000003 ,0);
\draw ( 2.8 ,-1/5) to ( 3.0 ,0);
\draw ( 3.0 ,-1/5) to ( 3.2 ,0);
\draw ( 3.2 ,-1/5) to ( 3.4000000000000004 ,0);
\draw ( 3.4 ,-1/5) to ( 3.6 ,0);
\draw ( 3.6 ,-1/5) to ( 3.8000000000000003 ,0);
\draw ( 3.8 ,-1/5) to ( 4.0 ,0);
\draw ( 4.0 ,-1/5) to ( 4.2 ,0);
\draw ( 4.2 ,-1/5) to ( 4.4 ,0);
\draw ( 4.4 ,-1/5) to ( 4.6000000000000005 ,0);
\draw ( 4.6 ,-1/5) to ( 4.8 ,0);
\draw ( 4.8 ,-1/5) to ( 5.0 ,0);
\draw ( 5.0 ,-1/5) to ( 5.2 ,0);
\draw ( 5.2 ,-1/5) to ( 5.4 ,0);
\draw ( 5.4 ,-1/5) to ( 5.6000000000000005 ,0);
\draw ( 5.6 ,-1/5) to ( 5.8 ,0);
\draw ( 5.8 ,-1/5) to ( 6.0 ,0);
\draw ( 6.0 ,-1/5) to ( 6.2 ,0);
\draw ( 6.2 ,-1/5) to ( 6.4 ,0);
\draw ( 6.4 ,-1/5) to ( 6.6000000000000005 ,0);
\draw ( 6.6 ,-1/5) to ( 6.8 ,0);
\draw ( 6.8 ,-1/5) to ( 7.0 ,0);
\draw ( 7.0 ,-1/5) to ( 7.2 ,0);
\draw ( 7.2 ,-1/5) to ( 7.4 ,0);
\draw (-8,0) to (-2,0);
\draw (1.6,0) to (7.4,0);

\draw [thick, ->] (-1.05,1) -- node[midway, above] {\Large $g(z,t)$ \normalsize} (0.25,1);

\draw (-5,0) node[circle,fill,inner sep=1.5pt]{} to[out=90,in=-90] (-4,1-0.2) to[out=90,in=-90] (-4.5,2-0.4) to[out=90,in=-155] (-4,3-0.4) node[circle,fill,inner sep=1.5pt]{}; 
\draw [very thick] (-4,3-0.4) node[circle,fill,inner sep=1.5pt]{} to[out=25,in=-175] (-3,3.3-0.4) node[circle,fill,inner sep=1.5pt]{};
\draw (-4,2.6) circle (0.5cm);
\node at (-4.3,3.3)  {$\pi$};
\node at (-3.5,2.0)  {$\pi$};

\draw [very thick] (5,0) node[circle,fill,inner sep=1.5pt]{} -- (5,1) node[circle,fill,inner sep=1.5pt]{};
\draw (5.5,0) arc (0:180:0.5cm);
\node at (5.5,0.5)  {$\frac{\pi}{2}$};
\node at (4.5,0.5)  {$\frac{\pi}{2}$};

\node at (-8,2.9)  {\LARGE $\mathbb {H}$ \normalsize}; 
\node at (2,2.9)  {\LARGE $\mathbb {H}$ \normalsize}; 

\end{tikzpicture}
\caption{}
    \label{fig3}
\end{figure}

The angle
$2\pi$ around $z_*$ shrinks to $\pi$ in the upper half plane around
$\xi(t)$. From this argument we can infer that $g(z,t)$ behaves like
$(\text{const.})\times\sqrt{z-z_*}$ in the neighborhood of $z_*$. It
follows that the angles $\pi$ on the both sides of the curve at the tip
of $\Gamma_t$ (the left picture in Figure \ref{fig3}) shrinks to
$\dfrac{\pi}{2}$ in the image (the right picture in Figure
\ref{fig3}). Thus the complement of $\Gamma_t$ in $\Gamma_{t+s}$ (the
bold curve in the left picture in Figure \ref{fig3}) is approximately
mapped to a vertical segment (the bold segment in the right picture in
Figure \ref{fig3}) by $g(z,t)$.

The map $g(z,t+s)$ is decomposed as $(g(z,t+s)\circ g(z,t)^{-1})\circ
g(z,t)$, where $g(z,t+s)\circ g(z,t)^{-1}$ is the map from the domain in
the right picture in Figure \ref{fig3} to the upper half plane and
approximated by the map in Figure \ref{fig2}, which justifies the
following approximation:
\[
    g(z,t+s)\circ g(z,t)^{-1}(w)
    \approx
    \xi(t) + \sqrt{(w -\xi(t))^2 +2c(t,s)}
    =
    w + \frac{c(t,s)}{w-\xi(t)} + O(w^{-2}),
\]
where $c(t,s)$ is determined by the length of the segment in the right
picuture in Figure \ref{fig3}. On the other hand, because of the
normalization \eqref{g(z,t)}, we can expand $g(z,t+s)\circ g(z,t)^{-1}$
as follows:
\[
    g(z,t+s)\circ g(z,t)^{-1}(w)
    =
    w + s w^{-1} + O(w^{-2}).
\]
Comparing the coefficients, we obtain $c(t,s)\approx s$.

Therefore, we can write, using (\ref{le1}), (\ref{le2}),
$$
g(z, t+s)\approx \xi(t)+\sqrt{(g(z, t) -\xi(t))^2 +2s}=
g(z,t)+\frac{s}{g(z, t)-\xi(t)} + O(s^2)
$$
which is equivalent to (\ref{le0}).

Another explanation (or the idea of the proof in \cite{d-mo-gum:13},
which follows the original idea by Kufarev et al., \cite{KSS68}; see
also \cite{take:14-1}) is as follows. The key ingredient is the {\em
Schwarz integral formula} in complex analysis, a corollary of the Cauchy
integral formula:
\begin{equation}
    f(z)
    =
    \frac{1}{\pi} \int_{\Real}
    \frac{\Im f(\xi)}{\xi-z} d\xi,
\label{schwarz}
\end{equation}
where $f:\HH \to \CC$ is a holomorphic function on the upper half plane
continuously extendable to $\bar\HH\to\CC$ and satisfies the following
estimate:
\begin{equation}
    \lim_{R\to\infty}\max_{\varphi\in[0,\pi]}|f(R e^{i\varphi})|=0.
\label{f(infty)=0}
\end{equation}
In fact, applying the Cauchy integral formula to the semi-circle
$[-R,R]\cup\{Re^{i\varphi}\mid \varphi\in[0,\pi]\}$, we have
\begin{align*}
    f(z) &=
    \frac{1}{2\pi i}\int_{-R}^R \frac{f(x)}{x-z} dx
    +
    \frac{1}{2\pi} \int_0^\pi 
    \frac{f(Re^{i\varphi})}{Re^{i\varphi}-z} R e^{i\varphi}\, d\varphi,
\\
    0 &=
    \frac{1}{2\pi i}\int_{-R}^R \frac{f(x)}{x-\bar z} dx
    +
    \frac{1}{2\pi} \int_0^\pi 
    \frac{f(Re^{i\varphi})}{Re^{i\varphi}-\bar z} R e^{i\varphi}\, 
    d\varphi.
\end{align*}
Taking the complex conjugate of the second equation and summing with the
first, we obtain
\[
    f(z) =
    \frac{1}{\pi} \int_{-R}^R \frac{\Im f(x)}{x-z} dx
    +
    \frac{1}{2\pi} \int_0^\pi 
    \frac{f(Re^{i\varphi})}{Re^{i\varphi}-z} R e^{i\varphi}\, d\varphi
    +
    \frac{1}{2\pi} \int_0^\pi 
    \frac{\overline{f(Re^{i\varphi})}}
         {Re^{-i\varphi}- z} R e^{-i\varphi}\, 
    d\varphi.
\]
Because of the estimate \eqref{f(infty)=0}, the last two terms converges
to $0$ when $R$ tends to infinity. Thus \eqref{schwarz} is proved.  (See
also Proposition 2.2 of \cite{d-mo-gum:13}.)

\medskip
Let us return to the chordal L\"owner equation. For $0\leq s \leq t$ we
define a map $h(z;s,t):\HH\to\HH$ by
\begin{equation}
    h(z;s,t) := g(g^{-1}(z,t),s).
\label{h(z;s,t)}
\end{equation}
(See Figure \ref{fig:h(z;s,t)}.)

\begin{figure}[h!]
\begin{tikzpicture}[scale=1]


\draw ( -8.0 ,-1/5) to ( -7.8 ,0);
\draw ( -7.8 ,-1/5) to ( -7.6 ,0);
\draw ( -7.6 ,-1/5) to ( -7.3999999999999995 ,0);
\draw ( -7.4 ,-1/5) to ( -7.2 ,0);
\draw ( -7.2 ,-1/5) to ( -7.0 ,0);
\draw ( -7.0 ,-1/5) to ( -6.8 ,0);
\draw ( -6.8 ,-1/5) to ( -6.6 ,0);
\draw ( -6.6 ,-1/5) to ( -6.3999999999999995 ,0);
\draw ( -6.4 ,-1/5) to ( -6.2 ,0);
\draw ( -6.2 ,-1/5) to ( -6.0 ,0);
\draw ( -6.0 ,-1/5) to ( -5.8 ,0);
\draw ( -5.8 ,-1/5) to ( -5.6 ,0);
\draw ( -5.6 ,-1/5) to ( -5.3999999999999995 ,0);
\draw ( -5.4 ,-1/5) to ( -5.2 ,0);
\draw ( -5.2 ,-1/5) to ( -5.0 ,0);
\draw ( -5.0 ,-1/5) to ( -4.8 ,0);
\draw ( -4.8 ,-1/5) to ( -4.6 ,0);
\draw ( -4.6 ,-1/5) to ( -4.3999999999999995 ,0);
\draw ( -4.4 ,-1/5) to ( -4.2 ,0);
\draw ( -4.2 ,-1/5) to ( -4.0 ,0);
\draw ( -4.0 ,-1/5) to ( -3.8 ,0);
\draw ( -3.8 ,-1/5) to ( -3.5999999999999996 ,0);
\draw ( -3.6 ,-1/5) to ( -3.4 ,0);
\draw ( -3.4 ,-1/5) to ( -3.1999999999999997 ,0);
\draw ( -3.2 ,-1/5) to ( -3.0 ,0);
\draw ( -3.0 ,-1/5) to ( -2.8 ,0);
\draw ( -2.8 ,-1/5) to ( -2.5999999999999996 ,0);
\draw ( -2.6 ,-1/5) to ( -2.4 ,0);
\draw ( -2.4 ,-1/5) to ( -2.1999999999999997 ,0);
\draw ( -2.2 ,-1/5) to ( -2.0 ,0);


\draw ( 1.6 ,-1/5) to ( 1.8 ,0);
\draw ( 1.8 ,-1/5) to ( 2.0 ,0);
\draw ( 2.0 ,-1/5) to ( 2.2 ,0);
\draw ( 2.2 ,-1/5) to ( 2.4000000000000004 ,0);
\draw ( 2.4 ,-1/5) to ( 2.6 ,0);
\draw ( 2.6 ,-1/5) to ( 2.8000000000000003 ,0);
\draw ( 2.8 ,-1/5) to ( 3.0 ,0);
\draw ( 3.0 ,-1/5) to ( 3.2 ,0);
\draw ( 3.2 ,-1/5) to ( 3.4000000000000004 ,0);
\draw ( 3.4 ,-1/5) to ( 3.6 ,0);
\draw ( 3.6 ,-1/5) to ( 3.8000000000000003 ,0);
\draw ( 3.8 ,-1/5) to ( 4.0 ,0);
\draw ( 4.0 ,-1/5) to ( 4.2 ,0);
\draw ( 4.2 ,-1/5) to ( 4.4 ,0);
\draw ( 4.4 ,-1/5) to ( 4.6000000000000005 ,0);
\draw ( 4.6 ,-1/5) to ( 4.8 ,0);
\draw ( 4.8 ,-1/5) to ( 5.0 ,0);
\draw ( 5.0 ,-1/5) to ( 5.2 ,0);
\draw ( 5.2 ,-1/5) to ( 5.4 ,0);
\draw ( 5.4 ,-1/5) to ( 5.6000000000000005 ,0);
\draw ( 5.6 ,-1/5) to ( 5.8 ,0);
\draw ( 5.8 ,-1/5) to ( 6.0 ,0);
\draw ( 6.0 ,-1/5) to ( 6.2 ,0);
\draw ( 6.2 ,-1/5) to ( 6.4 ,0);
\draw ( 6.4 ,-1/5) to ( 6.6000000000000005 ,0);
\draw ( 6.6 ,-1/5) to ( 6.8 ,0);
\draw ( 6.8 ,-1/5) to ( 7.0 ,0);
\draw ( 7.0 ,-1/5) to ( 7.2 ,0);
\draw ( 7.2 ,-1/5) to ( 7.4 ,0);
\draw ( 7.4 ,-1/5) to ( 7.6000000000000005 ,0);

\draw ( -3.0 ,-7.2) to ( -2.8 ,-7);
\draw ( -2.8 ,-7.2) to ( -2.6 ,-7);
\draw ( -2.6 ,-7.2) to ( -2.4 ,-7);
\draw ( -2.4 ,-7.2) to ( -2.2 ,-7);
\draw ( -2.2 ,-7.2) to ( -2.0 ,-7);
\draw ( -2.0 ,-7.2) to ( -1.8 ,-7);
\draw ( -1.8 ,-7.2) to ( -1.6 ,-7);
\draw ( -1.6 ,-7.2) to ( -1.4 ,-7);
\draw ( -1.4 ,-7.2) to ( -1.2 ,-7);
\draw ( -1.2 ,-7.2) to ( -1.0 ,-7);
\draw ( -1.0 ,-7.2) to ( -0.8 ,-7);
\draw ( -0.8 ,-7.2) to ( -0.6 ,-7);
\draw ( -0.6 ,-7.2) to ( -0.4 ,-7);
\draw ( -0.4 ,-7.2) to ( -0.2 ,-7);
\draw ( -0.2 ,-7.2) to ( 0 ,-7);
\draw ( 0 ,-7.2) to ( 0.2 ,-7);
\draw ( 0.2 ,-7.2) to ( 0.4 ,-7);
\draw ( 0.4 ,-7.2) to ( 0.6 ,-7);
\draw ( 0.6 ,-7.2) to ( 0.8 ,-7);
\draw ( 0.8 ,-7.2) to ( 1 ,-7);
\draw ( 1 ,-7.2) to ( 1.2 ,-7);
\draw ( 1.2 ,-7.2) to ( 1.4 ,-7);
\draw ( 1.4 ,-7.2) to ( 1.6 ,-7);
\draw ( 1.6 ,-7.2) to ( 1.8 ,-7);
\draw ( 1.8 ,-7.2) to ( 2 ,-7);
\draw ( 2 ,-7.2) to ( 2.2 ,-7);
\draw ( 2.2 ,-7.2) to (2.4 ,-7);
\draw ( 2.4 ,-7.2) to ( 2.6 ,-7);
\draw (-8,0) to (-2,0);
\draw (1.6,0) to (7.57,0);
\draw (-3,-7) to (2.6,-7);

\draw [thick, ->] (-1.25,1) -- node[midway, above] { $h(\xi;s,t)$ } (0.35,1);
\draw [thick, ->] (2,-4) to  (4,-2);
\draw [thick, ->] (-2,-4) to (-4,-2);
\node at (-4,-3) {{$g(z,t)$} };
\node at (4,-3) { {$g(z,s)$} };

\draw [->] (-5.85,0.65) to  (-4.15,0.65);
\draw [ <-] (-5.85,0.65)  -- node[midway, above] { $B_{s,t}$ } (-4.15,0.65) ;
\draw  (-6,0) to (-6,0.75);
\draw (-4,0) to (-4,0.75);

\draw[decorate,decoration=snake,segment amplitude=2.5pt,segment aspect=0.5,segment length=0.2cm](-7,0) -- (-6,0);
\draw[decorate,decoration=snake,segment amplitude=2.5pt,segment aspect=0.5,segment length=0.2cm](-4,0) -- (-3,0);
\draw [very thick] (-6,0) to (-4,0);
\node at (-5,0.01) [circle,fill,inner sep=1.5pt]{};
\node at (-5,-0.5) { {$\xi(t)$} };

\draw [very thick]  (5,0) node[circle,fill,inner sep=1.5pt]{}  to[out=90,in=-90] (4.5,1) to[out=90,in=-135] (5,1.5) node[circle,fill,inner sep=1.5pt]{}; 
\draw [decorate,decoration=snake,segment amplitude=2.5pt,segment aspect=0.5,segment length=0.2cm] (3.6,0) to (6,0);
\node at (5,-0.5) { {$\xi(s)$} };

\draw [decorate,decoration=snake,snake=coil,segment amplitude=2pt,segment aspect=0.5,segment length=0.12cm] 
(-1,-7) node[circle,fill,inner sep=1.5pt]{} to[out=90,in=-90] (0,-6) node[circle,fill,inner sep=1.5pt]{} ;
\node at (-1,-7.5)  { $\Gamma (0)$ };
\node at (0.8,-6)  { $\Gamma (s)$ };
\draw [very thick] (0,-6) to[out=90,in=-90] (-0.5,-5)  to[out=90,in=-135] (0,-4.5) node[circle,fill,inner sep=1.5pt,]{}; 
\node at (0.5,-4)  { $\Gamma (t)$ };

\end{tikzpicture}
\caption{}
\label{fig:h(z;s,t)}
    
\end{figure}


If we apply \eqref{schwarz} to $h(\xi;s,t)-\xi$, we obtain
\begin{equation}
    h(\xi;s,t)-\xi
    =
    \frac{1}{\pi}
    \int_{\Real} \frac{\Im(h(x;s,t)-x)}{x-\xi}\, dx.
    =
    \frac{1}{\pi}
    \int_{B_{s,t}} \frac{\Im(h(x;s,t))}{x-\xi}\, dx.
\label{h(z;s,t):integral-rep}
\end{equation}
since the integration variable $x$ is real and the imaginary part of
$h(z;s,t)$ is non-zero only on the bold interval $B_{s,t}$ in Figure
\ref{fig:h(z;s,t)}. Setting $\xi\mapsto g(z,t)$, we have
\begin{equation}
    g(z;s)-g(z,t)
    =
    \frac{1}{\pi}
    \int_{B_{s,t}} \frac{\Im(h(x;s,t))}{x-g(z,t)}\, dx.
\label{g(z,s)-g(z,t):chordal}
\end{equation}
On the other hand, setting $z=iy$ ($y>0$) in
\eqref{h(z;s,t):integral-rep} and taking the limit $y\to\infty$, we have
\begin{equation}
    t-s
    =
    \frac{1}{\pi}
    \int_{B_{s,t}} \Im(h(x;s,t))\, dx,
\label{t-s:integral-rep}
\end{equation}
since $h(\xi;s,t)=\xi+(s-t)\xi^{-1}+O(\xi^{-2})$. 

The ratio of \eqref{g(z,s)-g(z,t):chordal} and \eqref{t-s:integral-rep}
gives
\begin{equation}
    \frac{g(z,s)-g(z,t)}{s-t}
    =
    \frac{\int_{B_{s,t}} \frac{\Im h(x;s,t)}{x-g(z,t)} dx}
    {-\int_{B_{s,t}} \Im h(x;s,t) dx}
\label{g(z,s)-g(z,t)/s-t}
\end{equation}
for $0\leq s < t$.

When $s$ approaches to $t$ from below, $s\nearrow t$, the left hand side
of \eqref{g(z,s)-g(z,t)/s-t} converges to the left derivative of
$g(z,t)$ with respect to $t$. The segment $B_{s,t}$ shrinks to
$\xi(t)$ in this limit, which means that the right hand side of
\eqref{g(z,s)-g(z,t)/s-t} converges to
$\frac{1}{g(z,t)-\xi(t)}$. Thus we have an equation for
$h(\xi;s,t)$:
\begin{equation*}
    \frac{\der}{\der t}g(z,t)
    =
    \frac{1}{g(z,t)-\xi(t)},
\end{equation*}
where the differentiation is understood as the left derivative. We can
also prove the same differential equation with the right derivative, and
therefore the above equation holds as a usual differential equation.
namely, \eqref{le0}.

Below in section \ref{section:dkpreduction} we will derive the chordal L\"owner
equation in an absolutely different context of integrable hierarchies of 
nonlinear partial differential equations in the dispersionless limit as 
the consistency condition of the infinite dKP hierarchy with one-variable reduction. 

\subsection{Quadrant L\"owner equation}

Let us take the quadrant $Q=\{z\mid \Re z>0,\ \Im z>0 \}$ as the
reference domain for the chordal L\"owner equation instead of
$\HH$. Namely, we identify $\HH$ with $Q$ by a mapping of the form $h_a:
Q \owns z \mapsto \tilde z = z^2 -a(t) \in\HH$ ($a(t)\in\Real$) and
rewrite the equation \eqref{le0}. 

 If the function $g(z,t)$ of the form
\begin{equation}
    g(z,t) 
    = z + \sum_{n=1}^{\infty} v_n(t)z^{1-2n}
\label{g}
\end{equation}
 satisfies the
 equation
\begin{equation}
    \frac{\der g}{\der t}
    = - \frac{g}{g^2 - \xi^2} \frac{du}{dt}
\label{quadrant-loewner}
\end{equation}
 ($u=-v_1$), then $\tilde g(\tilde z,t):= g(\sqrt{\tilde z},t)^2 -
 2v_1(t)$ satisfies the hydrodynamic normalization condition
\begin{equation}
    \tilde g(\tilde z, t)
    =
    \tilde z +
    \tilde u(t) \tilde z^{-1} + O(\tilde z^{-2})
\label{gtilde:normalized}
\end{equation}
 ($\tilde u = 2 v_2 + v_1^2$) and the chordal L\"owner equation:
\begin{equation}
    \frac{\der \tilde g}{\der t}
    =
    \frac{1}{\tilde g - \tilde\xi} \frac{d\tilde u}{dt}.
\label{chordal-loewner}
\end{equation}
 Here the driving function $\tilde\xi=\tilde\xi(t)$ is 
\begin{equation}
    \tilde\xi(t) = \xi(t)^2-2v_1(t).
\label{tildexi=xi2-2v1}
\end{equation}
 Moreover, $\tilde u(t)$ satisfies
\begin{equation}
    \frac{d\tilde u}{dt} = - 2\xi^2  \frac{du}{dt}.
\label{dtildeu=-2xi2du}
\end{equation}

 Conversely, let $\tilde g(\tilde z,t) = \tilde z + \tilde u(t) \tilde
 z^{-1} + \ldots$ be a solution of the chordal L\"owner equation
 \eqref{chordal-loewner} with the driving function $\tilde\xi(t)$. Let
 $\xi(t)$ be a solution of the following ordinary differential
 equation\footnote{%
 The ordinary differential equation
 \eqref{V:diff-eq} for $\xi^2$ is a special case of Chini's equation
 (C.I.55 of \cite{kam:59}, p.303; $x\mapsto t$, $y\mapsto \xi(t)^2$,
 $n=-1$, $f(x)\mapsto d\tilde u/dt$, $g(x)\mapsto 0$, $h(x)\mapsto
 d\tilde\xi/dt$). It can be solved explicitly only in special cases.
 }:
\begin{equation}
    \frac{d}{dt} \xi^4 - 2 \frac{d\tilde\xi}{dt} \xi^2
    =
    2 \frac{d\tilde u}{dt}.
\label{V:diff-eq}
\end{equation}
 If we define $u(t)$ by
\begin{equation}
    u(t) = \frac{1}{2} (\tilde\xi(t) - \xi(t)^2),
\label{u=(tildexi-xi2)/2}
\end{equation}
 then $g(z,t) := \sqrt{\tilde g(z^2,t) - 2u(t)}$ is of
 the form \eqref{g} and satisfies equation
 \eqref{quadrant-loewner}. (Exactly speaking, we can choose a branch
 of the square root so that $g(z;t)$ is of the form \eqref{g} and
 satisfies \eqref{quadrant-loewner}.) 

The proof is a straightforward computation (see \cite{T13}, Proposition 5.1). 
\begin{figure}[h!]
\begin{tikzpicture}[scale=1]
\draw ( -5 , 3.0 ) to ( -5.0 , 3.0 );
\draw ( -5 , 3.6 ) to ( -4.4 , 3.0 );
\draw ( -5 , 4.2 ) to ( -3.8 , 3.0 );
\draw ( -5 , 4.8 ) to ( -3.2 , 3.0 );
\draw ( -5 , 5.4 ) to ( -2.6 , 3.0 );
\draw ( -5.0 , 6.0 ) to ( -2 , 3.0 );
\draw ( -4.4 , 6.0 ) to ( -2 , 3.6 );
\draw ( -3.8 , 6.0 ) to ( -2 , 4.2 );
\draw ( -3.2 , 6.0 ) to ( -2 , 4.8 );
\draw ( -2.6 , 6.0 ) to ( -2 , 5.4 );


\draw ( 1.6 , 3.0 ) to ( 1.6 , 3.0 );
\draw ( 1.6 , 3.6 ) to ( 2.2 , 3.0 );
\draw ( 1.6 , 4.2 ) to ( 2.8 , 3.0 );
\draw ( 1.6 , 4.8 ) to ( 3.4 , 3.0 );
\draw ( 1.6 , 5.4 ) to ( 4.0 , 3.0 );
\draw ( 1.6 , 6.0 ) to ( 4.6 , 3.0 );
\draw ( 2.2 , 6.0 ) to ( 4.6 , 3.6 );
\draw ( 2.8 , 6.0 ) to ( 4.6 , 4.2 );
\draw ( 3.4 , 6.0 ) to ( 4.6 , 4.8 );
\draw ( 4.0 , 6.0 ) to ( 4.6 , 5.4 );

\draw ( -8 , -3.6 ) to ( -8 , -3.6 );
\draw ( -8 , -3.0 ) to ( -7.4 , -3.6 );
\draw ( -8 , -2.4 ) to ( -6.8 , -3.6 );
\draw ( -8 , -1.8 ) to ( -6.2 , -3.6 );
\draw ( -8 , -1.2 ) to ( -5.6 , -3.6 );
\draw ( -8 , -0.6) to ( -5 , -3.6 );
\draw ( -8 , 0.0 ) to ( -4.4 , -3.6 );

\draw ( -7.4 , 0 ) to ( -3.8 , -3.6 );
\draw ( -6.8 , 0 ) to ( -3.2 , -3.6 );
\draw ( -6.2 , 0 ) to ( -2.6 , -3.6 );
\draw ( -5.6 , 0 ) to ( -2.0 , -3.6 );
\draw ( -5.0 , 0 ) to ( -2.0 , -3.0);
\draw ( -4.4 , 0 ) to ( -2.0 , -2.4);
\draw ( -3.8 , 0 ) to ( -2.0 , -1.8);
\draw ( -3.2 , 0 ) to ( -2.0 , -1.2);
\draw ( -2.6 , 0 ) to ( -2.0 , -0.6);

\draw ( 1.6, -3.6 ) to ( 1.6, -3.6 );
\draw ( 1.6, -3.0 ) to ( 2.2 , -3.6  );
\draw ( 1.6, -2.4 ) to ( 2.8 , -3.6  );
\draw ( 1.6, -1.8 ) to ( 3.4 , -3.6 );
\draw ( 1.6, -1.2 ) to ( 4.0 , -3.6 );
\draw ( 1.6, -0.6 ) to ( 4.6, -3.6 );
\draw ( 1.6, 0.0 ) to ( 5.2 , -3.6 );
\draw ( 2.2 , 0 ) to ( 5.8 , -3.6 );
\draw ( 2.8 , 0 ) to ( 6.4, -3.6);
\draw ( 3.4 , 0 ) to ( 7.0, -3.6 );
\draw ( 4 , 0 ) to ( 7.6 , -3.6 );
\draw ( 4.6 , 0 ) to ( 7.6 , -3.0);
\draw ( 5.2 , 0 ) to ( 7.6 , -2.4);
\draw ( 5.8 , 0 ) to ( 7.6 , -1.8);
\draw ( 6.4 , 0 ) to ( 7.6 , -1.2);
\draw ( 7 , 0 ) to ( 7.6 , -0.6);
\draw (-8,-3.6) to (-2,-3.6);
\draw (1.6,-3.6) to (7.6,-3.6);
\node at (-2.5,5.5) {\Large {$ z $} \normalsize};
\draw (-5,3) to (-2,3);
\draw (-5,3) to (-5,6);
\node at (4.1,5.5) {\Large {$ w $} \normalsize};
\draw (1.6,3) to (4.6,3);
\node at (-3,-1) {\Large {$ \tilde z $} \normalsize};
\draw (1.6,3) to (1.6,6);
\node at (6.6,-1.0) {\Large {$\tilde w $} \normalsize};

\draw [thick, ->] (-1.05,5) -- node[midway,above] { $g(z,t)=w$ } (0.25,5);
\draw [thick, ->] (-1.05,-2) -- node[midway, above] { $\tilde g(\tilde z,t)=\tilde w$ } (0.25,-2);

\draw [thick, ->] (-3.5,2.2) -- node[midway, right] { $\tilde z=z^2$ } (-3.5,1.1);
\draw [thick, ->] (3.1,1.1) -- node[midway, right] { $w=\sqrt {\tilde w -2u(t)}$ } (3.1,2.2);

\draw [very thick](-4,3)  to[out=120,in=-170](-3.5,4) to[out=0,in=170]  (-3,3.5) to[out=0,in=-90]  (-2.5,4.5) node[circle,fill,inner sep=1.5pt]{} ;

\draw [very thick](-4.6,-4.2+0.6)  to[out=110,in=-110] (-5.3,-3.2+0.6) to[out=80,in=30] (-6.3,-2.5+0.6) node[circle,fill,inner sep=1.5pt]{}; 

\draw [very thick] (2.3,3.01) -- (3.1,3.01) node[circle,fill,inner sep=1.6pt]{} -- (3.9,3.01);
\node at (3.1,2.51) {{$\xi (t)$}};

\draw [very thick] (4,-4.21+0.6) -- (5,-4.21+0.6) node[circle,fill,inner sep=1.6pt]{} -- (6,-4.21+0.6);
\node at (5,-4.11) {$\tilde \xi (t)$};
\end{tikzpicture}
\caption{}
    \label{quadrant}
\end{figure}


We see that a one-parameter family of conformal
mappings from $Q\setminus$(slit) to $Q$ with the normalization \eqref{g}
(the upper half of Figure \ref{quadrant}) satisfies the equation
\eqref{quadrant-loewner}, which we call the {\em quadrant L\"owner
equation}. Note that such a conformal map should be an odd function
because of the Schwarz reflection principle and therefore has the form
\eqref{g} up to normalization.

\bigskip
The quadrant L\"owner equation adds nothing essential to the L\"owner
theory, but is related to the dispersionless BKP hierarchy, as we shall
see later. In fact, the quadrant L\"owner equation was intorduced in
this context in \cite{T13}. 
(For another application see \cite{kat-kos:19}.)

\subsection{Radial L\"owner equation}

The radial version of the L\"owner equation differs from the chordal one
by a different normalization. 
Let $\Gamma_t$ be an arc of a smooth curve in the exterior of the unit disk
$\UU ^c$ starting from a point on the unit circle 
$\SSS$, and let $g(z, t)$ be the univalent conformal map
from $\UU^c \setminus \Gamma_t$ to $\UU^c$ normalized by the condition
$$
g(z,t)=e^{-\phi(t)}z +O(1), \quad z\to \infty , \quad t\in \RR . 
$$
According to the Riemann mapping theorem, such a map exists and is
unique.  The quantity $r=e^{\phi(t)}$ is called the conformal
radius of the domain $\UU^c \setminus \Gamma_t$.
As before we reparametrize the curve so that
$\phi(t)=t$:
$$
g(z,t)=e^{-t}z +O(1), \quad z\to \infty , \quad t\in \RR . 
$$
Note that a fixed interior point $z=\infty$ of $\UU^c$ is mapped to a
fixed inner point $\infty$, while in the chordal case a fixed boundary
point $z=\infty$ is mapped to a fixed boundary point $\infty$. This
difference in normalization is the essential difference between the two
types of the L\"owner equation.

It appears that there exists a 
continuous real-valued function $\xi(t)$ 
(called the driving function) such that 
$g(z,t)$ satisfies the differential equation
\beq\label{le3}
\frac{\p g(z, t)}{\p t}=-g(z,t)\,
\frac{g(z, t)+\eta (t)}{g(z, t)-\eta (t)}, \qquad g(z, 0)=z, \quad \eta (t)=e^{i\xi(t)}.
\eeq
It is now called the radial L\"owner equation to distinguish it from the chordal case.
This equation was found by L\"owner in 1923 \cite{L1923}. It is a powerful tool in
geometric function theory, for example, to evaluate coefficients of univalent functions. 
One of the examples, to which the radial L\"owner equation was successfully applied, is the 
famous Bieberbach conjecture. 

For example, 
if $\Gamma_t = [\eta , \eta \rho]$ is the straight segment from $\eta \in \SSS$ to
the point $\eta \rho$ (orthogonal to the unit circle), where 
$$
\rho = 2e^t -1+2\sqrt{e^{2t}-e^t} \, \in \RR ,
$$
then $\eta(t)=\eta =\mbox{const}$ and
\beq\label{le4}
g(z,t)=-\eta +\frac{1}{2}\, e^{-t}(z+\eta)\left (1+\frac{\eta}{z}+\frac{1}{z}
\sqrt{z^2 +2\eta z(1-2e^t)+\eta^2}\, \right ).
\eeq
Note that $g(\eta \rho , t)=\eta$ and that 
if $t\to 0$ and $|z-\eta |$ is bounded from below, then
\beq\label{le5}
g(z, t)=z-z \frac{z+\eta}{z-\eta}\, t +O(t^2).
\eeq

We can explain (\ref{le3}) in the same way as in the chordal case.

\begin{figure}[h!]
\begin{tikzpicture}[scale=1]
\node at (-8,3)  {\LARGE $\mathbb {U}^{C}$ \normalsize}; 
\node at (2,3)  {\LARGE $\mathbb {U}^{C}$ \normalsize}; 

\draw [thick, ->] (-1.45,1) -- node[midway, above] {\Large $g(z,t)$ \normalsize} (-0.15,1);


\draw [very thick] ( -6.15147186258 , 1.34852813742 )  node[circle,fill,inner sep=1.5pt]{} -- ( -5.30294372515 , 2.19705627485 ) node[circle,fill,inner sep=1.5pt]{} ;
\node[circle,fill,inner sep=1.5pt] at ( -7 , 0.5 ) {};
\node[circle,fill,inner sep=1.5pt] at ( 3 , 0.5 ) {};
\draw ( -7 , 0.5 ) circle ( 1.2 cm);
\draw ( 3 , 0.5 ) circle ( 1.2 cm);
\draw ( -5.8 , 0.5 ) -- ( -5.96180339887 , 0.617557050458 );
\draw ( -5.81477399129 , 0.312278641952 ) -- ( -5.95619534752 , 0.453699998189 );
\draw ( -5.85873218045 , 0.12917960675 ) -- ( -5.9762892309 , 0.290983005625 );
\draw ( -5.93079217097 , -0.0447885996875 ) -- ( -6.02159027092 , 0.13341270515 );
\draw ( -6.02917960675 , -0.205342302751 ) -- ( -6.09098300563 , -0.0151309994919 );
\draw ( -6.15147186258 , -0.348528137424 ) -- ( -6.18275875558 , -0.150990469305 );
\draw ( -6.29465769725 , -0.47082039325 ) -- ( -6.29465769725 , -0.27082039325 );
\draw ( -6.45521140031 , -0.569207829026 ) -- ( -6.4239245073 , -0.371670160907 );
\draw ( -6.62917960675 , -0.641267819554 ) -- ( -6.56737620788 , -0.451056516295 );
\draw ( -6.81227864195 , -0.685226008714 ) -- ( -6.721480542 , -0.507024703876 );
\draw ( -7.0 , -0.7 ) -- ( -6.88244294954 , -0.538196601125 );
\draw ( -7.18772135805 , -0.685226008714 ) -- ( -7.04630000181 , -0.543804652477 );
\draw ( -7.37082039325 , -0.641267819554 ) -- ( -7.20901699437 , -0.523710769096 );
\draw ( -7.54478859969 , -0.569207829026 ) -- ( -7.36658729485 , -0.478409729078 );
\draw ( -7.70534230275 , -0.47082039325 ) -- ( -7.51513099949 , -0.409016994375 );
\draw ( -7.84852813742 , -0.348528137424 ) -- ( -7.6509904693 , -0.317241244416 );
\draw ( -7.97082039325 , -0.205342302751 ) -- ( -7.77082039325 , -0.205342302751 );
\draw ( -8.06920782903 , -0.0447885996875 ) -- ( -7.87167016091 , -0.0760754926955 );
\draw ( -8.14126781955 , 0.12917960675 ) -- ( -7.9510565163 , 0.0673762078751 );
\draw ( -8.18522600871 , 0.312278641952 ) -- ( -8.00702470388 , 0.221480542004 );
\draw ( -8.2 , 0.5 ) -- ( -8.03819660113 , 0.382442949542 );
\draw ( -8.18522600871 , 0.687721358048 ) -- ( -8.04380465248 , 0.546300001811 );
\draw ( -8.14126781955 , 0.87082039325 ) -- ( -8.0237107691 , 0.709016994375 );
\draw ( -8.06920782903 , 1.04478859969 ) -- ( -7.97840972908 , 0.86658729485 );
\draw ( -7.97082039325 , 1.20534230275 ) -- ( -7.90901699437 , 1.01513099949 );
\draw ( -7.84852813742 , 1.34852813742 ) -- ( -7.81724124442 , 1.1509904693 );
\draw ( -7.70534230275 , 1.47082039325 ) -- ( -7.70534230275 , 1.27082039325 );
\draw ( -7.54478859969 , 1.56920782903 ) -- ( -7.5760754927 , 1.37167016091 );
\draw ( -7.37082039325 , 1.64126781955 ) -- ( -7.43262379212 , 1.4510565163 );
\draw ( -7.18772135805 , 1.68522600871 ) -- ( -7.278519458 , 1.50702470388 );
\draw ( -7.0 , 1.7 ) -- ( -7.11755705046 , 1.53819660113 );
\draw ( -6.81227864195 , 1.68522600871 ) -- ( -6.95369999819 , 1.54380465248 );
\draw ( -6.62917960675 , 1.64126781955 ) -- ( -6.79098300563 , 1.5237107691 );
\draw ( -6.45521140031 , 1.56920782903 ) -- ( -6.63341270515 , 1.47840972908 );
\draw ( -6.29465769725 , 1.47082039325 ) -- ( -6.48486900051 , 1.40901699437 );
\draw ( -6.15147186258 , 1.34852813742 ) -- ( -6.3490095307 , 1.31724124442 );
\draw ( -6.02917960675 , 1.20534230275 ) -- ( -6.22917960675 , 1.20534230275 );
\draw ( -5.93079217097 , 1.04478859969 ) -- ( -6.12832983909 , 1.0760754927 );
\draw ( -5.85873218045 , 0.87082039325 ) -- ( -6.0489434837 , 0.932623792125 );
\draw ( -5.81477399129 , 0.687721358048 ) -- ( -5.99297529612 , 0.778519457996 );
\draw ( 4.2 , 0.5 ) -- ( 4.03819660113 , 0.617557050458 );
\draw ( 4.18522600871 , 0.312278641952 ) -- ( 4.04380465248 , 0.453699998189 );
\draw ( 4.14126781955 , 0.12917960675 ) -- ( 4.0237107691 , 0.290983005625 );
\draw ( 4.06920782903 , -0.0447885996875 ) -- ( 3.97840972908 , 0.13341270515 );
\draw ( 3.97082039325 , -0.205342302751 ) -- ( 3.90901699437 , -0.0151309994919 );
\draw ( 3.84852813742 , -0.348528137424 ) -- ( 3.81724124442 , -0.150990469305 );
\draw ( 3.70534230275 , -0.47082039325 ) -- ( 3.70534230275 , -0.27082039325 );
\draw ( 3.54478859969 , -0.569207829026 ) -- ( 3.5760754927 , -0.371670160907 );
\draw ( 3.37082039325 , -0.641267819554 ) -- ( 3.43262379212 , -0.451056516295 );
\draw ( 3.18772135805 , -0.685226008714 ) -- ( 3.278519458 , -0.507024703876 );
\draw ( 3.0 , -0.7 ) -- ( 3.11755705046 , -0.538196601125 );
\draw ( 2.81227864195 , -0.685226008714 ) -- ( 2.95369999819 , -0.543804652477 );
\draw ( 2.62917960675 , -0.641267819554 ) -- ( 2.79098300563 , -0.523710769096 );
\draw ( 2.45521140031 , -0.569207829026 ) -- ( 2.63341270515 , -0.478409729078 );
\draw ( 2.29465769725 , -0.47082039325 ) -- ( 2.48486900051 , -0.409016994375 );
\draw ( 2.15147186258 , -0.348528137424 ) -- ( 2.3490095307 , -0.317241244416 );
\draw ( 2.02917960675 , -0.205342302751 ) -- ( 2.22917960675 , -0.205342302751 );
\draw ( 1.93079217097 , -0.0447885996875 ) -- ( 2.12832983909 , -0.0760754926955 );
\draw ( 1.85873218045 , 0.12917960675 ) -- ( 2.0489434837 , 0.0673762078751 );
\draw ( 1.81477399129 , 0.312278641952 ) -- ( 1.99297529612 , 0.221480542004 );
\draw ( 1.8 , 0.5 ) -- ( 1.96180339887 , 0.382442949542 );
\draw ( 1.81477399129 , 0.687721358048 ) -- ( 1.95619534752 , 0.546300001811 );
\draw ( 1.85873218045 , 0.87082039325 ) -- ( 1.9762892309 , 0.709016994375 );
\draw ( 1.93079217097 , 1.04478859969 ) -- ( 2.02159027092 , 0.86658729485 );
\draw ( 2.02917960675 , 1.20534230275 ) -- ( 2.09098300563 , 1.01513099949 );
\draw ( 2.15147186258 , 1.34852813742 ) -- ( 2.18275875558 , 1.1509904693 );
\draw ( 2.29465769725 , 1.47082039325 ) -- ( 2.29465769725 , 1.27082039325 );
\draw ( 2.45521140031 , 1.56920782903 ) -- ( 2.4239245073 , 1.37167016091 );
\draw ( 2.62917960675 , 1.64126781955 ) -- ( 2.56737620788 , 1.4510565163 );
\draw ( 2.81227864195 , 1.68522600871 ) -- ( 2.721480542 , 1.50702470388 );
\draw ( 3.0 , 1.7 ) -- ( 2.88244294954 , 1.53819660113 );
\draw ( 3.18772135805 , 1.68522600871 ) -- ( 3.04630000181 , 1.54380465248 );
\draw ( 3.37082039325 , 1.64126781955 ) -- ( 3.20901699437 , 1.5237107691 );
\draw ( 3.54478859969 , 1.56920782903 ) -- ( 3.36658729485 , 1.47840972908 );
\draw ( 3.70534230275 , 1.47082039325 ) -- ( 3.51513099949 , 1.40901699437 );
\draw ( 3.84852813742 , 1.34852813742 ) -- ( 3.6509904693 , 1.31724124442 );
\draw ( 3.97082039325 , 1.20534230275 ) -- ( 3.77082039325 , 1.20534230275 );
\draw ( 4.06920782903 , 1.04478859969 ) -- ( 3.87167016091 , 1.0760754927 );
\draw ( 4.14126781955 , 0.87082039325 ) -- ( 3.9510565163 , 0.932623792125 );
\draw ( 4.18522600871 , 0.687721358048 ) -- ( 4.00702470388 , 0.778519457996 );
\node at (-5.8,1.33)  {\Large $\eta$ \normalsize};
\node at (-4.7,2.3)  {\Large $\eta\rho$ \normalsize};

\draw [very thick] (3+1.14126781955,0.5+0.37082039325) node[circle,fill,inner sep=1.5pt]{} to[out=108,in=-45] (3+0.848528137424,0.5+0.848528137424) node[circle,fill,inner sep=1.5pt]{} to[out=135,in=-12] (3+0.37082039325,0.5+1.14126781955) node[circle,fill,inner sep=1.5pt]{};
\node at (3+1+1.1,0.5+1.1)  {\Large $g(\eta\rho)=\eta$ \normalsize};

\end{tikzpicture}
\caption{}
    \label{fig5}
\end{figure}
\bigskip
%

Let the curve evolve for a ``time'' $t$ and then for a further short 
``time'' $s$, $s\to 0$. The image of $\UU^c \setminus 
\Gamma_{t+s}$ under $g(z,t)$ is 
$\UU^c$ with a cut which is 
a short segment orthogonal to the unit circle starting from the point
$\eta (t)$ on $\SSS$. Therefore, we can write, using (\ref{le4}), (\ref{le5}),
$$
g(z, t+s)\approx 
g(z,t)-g(z,t) \frac{g(z, t)+\eta(t)}{g(z, t)-\eta(t)}\, s +O(s^2)
$$
which is equivalent to (\ref{le3}). 

In another explanation (the idea of the proof in, e.g., \cite{Duren},
\cite{kom:40}, \cite{kom:44} \S34) the {\em complex Poincar\'e integral
formula} 
\begin{equation}
    f(z) 
    = 
    \frac{1}{2\pi} \int_0^{2\pi}
    \Re f(e^{i\theta}) \frac{e^{i\theta}+z}{e^{i\theta}-z} d\theta
    + i\Im f(0)
\label{poincare}
\end{equation}
for a holomorphic function $f(z)$ on the unit disk $\UU$ plays the role
of the Schwarz integral formula \eqref{schwarz} in the chordal
case. 

Let us consider a family of conformal mappings $f(z,t)$ from
$\UU\setminus \Gamma((0,t])$ to $\UU$. Here $\Gamma:[0,+\infty)\to
\bar\UU$ is a Jordan curve with the condition $|\Gamma(0)|=1$ and
$f(z,t)$ is normalized by $f(z,t)=e^t z + O(z^2)$. We derive the
differential equation
\begin{equation}
    \frac{\der f(z,t)}{\der t}
    =
    f(z,t)\frac{\lambda(t)+f(z,t)}{\lambda(t)-f(z,t)},
\label{radial}
\end{equation}
from which equation \eqref{le3} follows immediately
($g(z,t)=1/f(z^{-1},t)$, $\eta(t)=1/\lambda(t)$).

As in the chordal case (cf.\ \eqref{h(z;s,t)}), we
define a map $h(z;s,t)$ for $z\in\UU$ ($0<s<t$) by
\begin{equation}
    h(z;s,t) := f(f^{-1}(z,t),s)
    = e^{s-t}z+O(z^2).
\label{h(z;s,t):radial}
\end{equation}
In Figure 7 the image of the tip of the curve by
$f(z,t)$ is denoted by $\lambda(t)$: $\lambda(t)= f(\Gamma(t),t)$.

\begin{figure}[h!]
\begin{tikzpicture}[scale=0.8]

\draw (-5,0) circle (2.5cm);
\draw (5,0) circle (3cm);
\draw (0,-8) circle (3.5cm);

\draw[->]  (-2,0) -- node[midway, above, sloped] {$h(z;s,t)$} (1.5,0);
\draw[->]  (-3,-5) -- node[midway, above, sloped] {$f(z,t)$} (-4,-3);
\draw[->]  (3,-5) -- node[midway, above, sloped] {$f(z,s)$} (4,-3);

\fill[black] (-5+2.5,0) circle (0.1cm);
\node at (-3,0){$1$};

\fill[black] (5+3,0) circle (0.1cm);
\node at (7.5,0){$1$};

\fill[black] (3.5,-8) circle (0.1cm);
\node at (3,-8){$1$};

\draw[line width=0.6mm] ( -3.084888892202555 , 1.6069690242163481 ) arc (40:80:2.5);
\draw[decorate,decoration=snake,snake=coil,segment amplitude=2pt,segment aspect=0.5,segment length=0.12cm] (-2.538,0.4348 ) arc (10:40:2.5);
\draw[decorate,decoration=snake,snake=coil,segment amplitude=2pt,segment aspect=0.5,segment length=0.12cm] ( -4.565879555832674 , 2.46201938253052 ) arc (80:110:2.5);
\draw ( -3.238097780826351 , 1.47841150227904 ) -- ( -2.931680003578759 , 1.7355265461536562 );
\draw ( -4.5311499202992875 , 2.6589809331329617 ) -- ( -4.60060919136606 , 2.265057831928078 );

\fill[black] ( -3.75 , 2.1650635094610964 ) circle (0.1cm);
\node at ( -3.75+0.6 , 2.1650635094610964 +0.6 ){$\lambda(t)$};
\draw[decorate,decoration=snake,snake=coil,segment amplitude=2pt,segment aspect=0.5,segment length=0.12cm] ( ( 7.598076211353316 , 1.4999999999999998 ) arc (30:90:3);

\fill[black] ( 6.5 , 2.598076211353316 ) circle (0.1cm);
\node at ( 6.5+0.6 , 2.598076211353316+0.6 ){$\lambda(s)$};

\draw[line width=0.6mm]  ( 6.5 , 2.598076211353316 ) to[out=-110,in=0] ( 5 , 2) to[out=-180,in=45] ( 4 , 1);

\fill[black] ( 2.4748737341529163 , -5.525126265847084 ) circle (0.1cm);
\node at ( 2.4748737341529163 +1, -5.525126265847084 ){$\Gamma(0)$};
\draw[decorate,decoration=snake,snake=coil,segment amplitude=2pt,segment aspect=0.5,segment length=0.12cm] 
( 2.4748737341529163 , -5.525126265847084 ) to[out=-130,in=-40] (0.3,-6);

\draw[line width=0.6mm]  (0.3,-6) to[out=140,in=90] ( -1.5, -7.5); 
\fill[black] (0.3,-6) circle (0.1cm);
\node at (0.3,-6+0.6){$\Gamma(s)$};

\fill[black] ( -1.5, -7.5) circle (0.1cm);
\node at ( -1.5-0.6, -7.5-0.7){$\Gamma(t)$};

\end{tikzpicture}
\label{fig:h(z;s,t):radial}
\caption{}
\end{figure}

Applying \eqref{poincare} to the function $\log(h(z;s,t)/z)$,
we have
\begin{equation}
    \log\frac{h(z;s,t)}{z}
    =
    \frac{1}{2\pi} \int_0^{2\pi}
    \log|h(e^{i\theta};s,t)|\,
    \frac{e^{i\theta}+z}{e^{i\theta}-z} d\theta,
\label{log(h/z):radial}
\end{equation}
and, putting $z\mapsto f(z,t)$,
\begin{equation}
    \log\frac{f(z,s)}{f(z,t)}
    =
    \frac{1}{2\pi i} \int_0^{2\pi}
    \log|h(e^{i\theta};s,t)|\,
    \frac{e^{i\theta}+f(z,t)}{e^{i\theta}-f(z,t)} d\theta,
\label{log(f/f)}
\end{equation}
which corresponds to \eqref{g(z,s)-g(z,t):chordal} in the chordal case.

On the other hand, putting $z=0$ in \eqref{log(h/z):radial}, we have
\begin{equation}
    s-t
    =
    \frac{1}{2\pi}\int_0^{2\pi}
    \log|h(e^{i\theta};s,t)|\, d\theta,
\label{s-t=integral:radial}
\end{equation}
which substitutes \eqref{t-s:integral-rep} in the chordal case.

Note that $\log|h(e^{i\theta};s,t)|=0$ if $e^{i\theta}$ is mapped to the
boundary of $\UU$ by $h(z;s,t)$, namely, if it does not lie on the bold
arc of the left upper circle in Figure \ref{fig:h(z;s,t):radial}. Since
the bold arc shrinks to $\lambda(t)$, when $s\nearrow t$, the ratio of
\eqref{log(f/f)} and \eqref{s-t=integral:radial} converges to
\begin{equation}
    \frac{\der \log f(z,t)}{\der t}
    =
    \frac{\lambda(t)+f(z,t)}{\lambda(t)-f(z,t)},
\label{radial-loewner:temp}
\end{equation}
where the differentiation in the left hand side is understood as the
left derivative. We can also prove the same differential equation with
the right derivative, and therefore the above equation holds as a usual
differential equation. Thus we have obtained \eqref{radial}.

Below in section \ref{section:dtodareduction} we will derive the radial
L\"owner equation in the context of integrable hierarchies as the
consistency condition of the infinite dToda hierarchy with one-variable
reduction.

\label{section:lowner}

\subsection{Komatu-L\"owner equation}
\label{subsec:komatu-loewner}

The next example of the L\"owner-type equations is an equation for the
doubly connected domain found by Komatu (\cite{Komatu}, \cite{kom:49}
\S84). (See also
\cite{Goluzin,Alexandrov,CDMG1,CDMG2,review,fuk-kan:14}\footnote{We
partly follow the normalizations of maps and annuli in \cite{Goluzin}
and partly those in \cite{fuk-kan:14}, so that the resulting
differential equation has the same form as the reduction of the
dispersionless DKP hierarchy in \S\ref{subsec:1-var-reduction:dDKP}.}.)

Let $Q$ be a positive real number, $0<Q<1$, $\Ann_Q$ be an annulus,
$\{z\mid 1<|z|<Q^{-1}\}$ and $\Gamma$ be a Jordan curve in the closure
of $\Ann_Q$, $\Gamma:[0,+\infty)\to\Ann_Q$. We assume that $\Gamma$
starts from the outer boundary of $\Ann_Q$, $|\Gamma(0)|=Q^{-1}$, and
the other part lies completely in $\Ann_Q$, $\Gamma((0,+\infty))\subset
\Ann_Q$. Then, as is known in the geometric function theory, we can
reparametrize the curve, so that there exists a unique conformal map
$g(z,q)$ for each $q\in[0,+\infty)$ which maps
$\Ann_Q\setminus\Gamma((0,q])$ onto $\Ann_q=\{w\mid 1<|w|<q^{-1}\}$ with
the normalization condition,
\begin{equation}
   g(1,q)=1.
\label{g(1,1)=1}
\end{equation}
Let $\Lambda(q)$ be the image of $\Gamma(q)$ by $g(z,q)$,
\begin{equation}
    \Lambda(q):=g(\Gamma(q),q).
\label{Lambda(q)}
\end{equation}

\begin{figure}[h!]
\begin{tikzpicture}[scale=1]

\draw [->] (-0.5,0) -- node[midway, above] { $g(z,q )$ } (1,0);

\draw (-5,0) circle (1cm);
\draw (5,0) circle (1cm);

\draw (5,0) circle (2.5cm);
\draw (-5,0) circle (3cm);


\node at (-5,-1.6){ ${\mathbb A}_q$};

\node at (-4.5,0) { $1$};
\node[circle,fill,inner sep=1.5pt] at ( -2 , 0 ) {};
\node[circle,fill,inner sep=1.5pt] at ( -4 , 0 ) {};
\node at ( -2-0.75 , 0 ) { $1/Q$ };

\node[circle,fill,inner sep=1.5pt] at ( -3,2.23) {};
\draw  [very thick] (-3,2.23)  to[out=170,in=-110] (-4,2) to[out=80,in=30] (-5,1.55) node[circle,fill,inner sep=1.5pt]{}; 


\node[circle,fill,inner sep=1.5pt] at ( 6 , 0 ) {};
\node[circle,fill,inner sep=1.5pt] at ( 7.5 , 0 ) {};
\node at ( 5.5 , 0)   { $1$};
\node at ( 7, 0 )  { $1/q$};

\draw [very thick] (7.462, 0.434) arc (10:100:2.5cm);
\node[circle,fill,inner sep=1.5pt] at ( 6.6,1.915) {};

\end{tikzpicture}
\caption{}
\label{fig:g-q}
\end{figure}

Then $g(z,q)$ satisfies the following {\em Komatu-L\"owner
equation}\footnote{Sometimes it is called the {\em Goluzin-Komatu
equation} (e.g., in \cite{CDMG1}, \cite{akh-zab:14-2}, \cite{ATZ17}).}.
\begin{equation}
 \begin{split}
    \frac{\der \log g(z,q)}{\der \log q}
    &=
    \calK_q(g(z,q),\Lambda(q)) - \calK_q(1,\Lambda(q))
\\
    &=
    -\frac{1}{\pi i}
    \left(
     \zeta_1\left(
             \frac{\log g(z,q) - \log\Lambda(q)}{2\pi i},
             \frac{\log q}{\pi i}
            \right)
     +
     \zeta_1\left(
             \frac{\log\Lambda(q)}{2\pi i},
             \frac{\log q}{\pi i}
            \right)
    \right)
 \end{split}
\label{komatu-loewner}
\end{equation}
The functions $\calK_q(z,\zeta)$ (Villat's kernel) and $\zeta_1(u,\tau)$
are defined by \eqref{villat-kernel} and \eqref{zeta-wp}
respectively. The expression with $\zeta_1$ obtained from the expression
with $\calK_q$ by the relations \eqref{zeta1:expansion} and
$\zeta_1(-u)=-\zeta_1(u)$.

The idea of the proof is essentially the same as in the chordal L\"owner
case. The substitute of the Schwarz integral formula \eqref{schwarz} is
{\em Villat's formula}: Let $f: \bar\Ann_q\to\Comp$ be a continuous
function, which is holomorphic on $\Ann_q$. Then $f(z)$ is expressed as
follows: 
\begin{equation}
 \begin{split}
    f(z) ={}&
    \frac{1}{2\pi}
    \int_0^{2\pi} 
     \Re f(q^{-1}e^{i\theta})\, \calK_q(z,q^{-1}e^{i\theta})\,
    d\theta
    -
    \frac{1}{2\pi}
    \int_0^{2\pi} 
     \Re f(e^{i\theta})\, \calK_q(z,e^{i\theta})\,
    d\theta
\\
    &-
    \frac{1}{2\pi}
    \int_0^{2\pi} \Re f(e^{i\theta})\, d\theta
    +
    \frac{1}{2\pi}
    \int_0^{2\pi} \Im f(e^{i\theta})\, d\theta.
 \end{split}
\label{villat}
\end{equation}
Furthermore,
\begin{equation}
    \int_0^{2\pi} \Re f(e^{i\theta})\, d\theta
    =
    \int_0^{2\pi} \Re f(e^{i\theta})\, d\theta.
\label{villat-boundary}
\end{equation}
Assume that $\Re f(z) =: A \in\Real$ is constant on the inner boundary
$\der_{\text{inner}}\Ann_q=\{z\mid |z|=1\}$ of $\Ann_q$.  Villat's
formula gives the following expression of $f(z)$:
\begin{equation}
    f(z)
    =
    \frac{1}{2\pi}
    \int_0^{2\pi} \Re f(e^{i\theta})\, \calK_q(z,e^{i\theta})\, d\theta
    + ic,
\label{villat-corollary}
\end{equation}
where $c$ is a real constant. Moreover the constant $A$ is expressed by
the following integral:
\begin{equation}
    A
    =
    \frac{1}{2\pi}
    \int_0^{2\pi} \Re f(q^{-1}e^{i\theta})\, d\theta.
\label{villat-suppl}
\end{equation}

As in the chordal case (cf.\ \eqref{h(z;s,t)}), we
define a map $h(\xi;q^*,q)$ for $\xi\in\Ann_q$ ($0<q^*<q<1$) by
\begin{equation}
    h(\xi;q^*,q) := g(g^{-1}(\xi,q),q^*).
\label{h(xi;q*,q)}
\end{equation}
(See Figure \ref{fig:h(xi;q*,q)}.)

\begin{figure}[h!]
\begin{tikzpicture}[scale=1]

\draw (-5,0) circle (1cm);
\draw (5,0) circle (1cm);
\draw (0,-8) circle (1cm);

\draw (-5,0) circle (2.5cm);
\draw (5,0) circle (3cm);
\draw (0,-8) circle (3.5cm);

\draw[->]  (-2,0) -- node[midway, above, sloped] {$h(\xi;q^*,q)$} (1.5,0);
\draw[->]  (-3,-5) -- node[midway, above, sloped] {$g(z,q)$} (-4,-3);
\draw[->]  (3,-5) -- node[midway, above, sloped] {$g(z,q^*)$} (4,-3);

\fill[black] (-5+1,0) circle (0.1cm);
\node at (-4.5,0){$1$};
\fill[black] (-5+2.5,0) circle (0.1cm);
\node at (-3,0){$\frac{1}{q}$};
\fill[black] (5+1,0) circle (0.1cm);
\node at (5.5,0){$1$};
\fill[black] (5+3,0) circle (0.1cm);
\node at (7.5,0){$\frac{1}{q^*}$};
\fill[black] (1,-8) circle (0.1cm);
\node at (0.5,-8){$1$};
\fill[black] (3.5,-8) circle (0.1cm);
\node at (3,-8){$\frac{1}{Q}$};

\draw[line width=0.6mm] ( -3.084888892202555 , 1.6069690242163481 ) arc (40:80:2.5);
\draw[decorate,decoration=snake,snake=coil,segment amplitude=2pt,segment aspect=0.5,segment length=0.12cm] (-2.538,0.434) arc (10:40:2.5);
\draw[decorate,decoration=snake,snake=coil,segment amplitude=2pt,segment aspect=0.5,segment length=0.12cm] ( -4.565879555832674 , 2.46201938253052 ) arc (80:110:2.5);
\draw ( -3.238097780826351 , 1.47841150227904 ) -- ( -2.931680003578759 , 1.7355265461536562 );
\draw ( -4.5311499202992875 , 2.6589809331329617 ) -- ( -4.60060919136606 , 2.265057831928078 );

\fill[black] ( -3.75 , 2.1650635094610964 ) circle (0.1cm);
\node at ( -3.75+0.6 , 2.1650635094610964 +0.6 ){$\Lambda(q)$};
\draw[decorate,decoration=snake,snake=coil,segment amplitude=2pt,segment aspect=0.5,segment length=0.12cm] ( ( 7.598076211353316 , 1.4999999999999998 ) arc (30:90:3);

\fill[black] ( 6.5 , 2.598076211353316 ) circle (0.1cm);
\node at ( 6.5+0.6 , 2.598076211353316+0.6 ){$\Lambda(q^*)$};

\draw[line width=0.6mm]  ( 6.5 , 2.598076211353316 ) to[out=-110,in=0] ( 5 , 2) to[out=-180,in=45] ( 4 , 1);

\fill[black] ( 2.4748737341529163 , -5.525126265847084 ) circle (0.1cm);
\node at ( 2.4748737341529163 +1, -5.525126265847084 ){$\Gamma(0)$};
\draw[decorate,decoration=snake,snake=coil,segment amplitude=2pt,segment aspect=0.5,segment length=0.12cm] 
( 2.4748737341529163 , -5.525126265847084 ) to[out=-130,in=-40] (0.3,-6);

\draw[line width=0.6mm]  (0.3,-6) to[out=140,in=90] ( -1.5, -7.5); 
\fill[black] (0.3,-6) circle (0.1cm);
\node at (0.3,-6+0.6){$\Gamma(q^*)$};

\fill[black] ( -1.5, -7.5) circle (0.1cm);
\node at ( -1.5-0.6, -7.5-0.7){$\Gamma(q)$};

\node at (-5,-1.6){${\mathbb A}_q$};

\end{tikzpicture}
\caption{}
\label{fig:h(xi;q*,q)}
\end{figure}

Applying \eqref{villat-corollary} to
the function 
\begin{equation}
    \Phi(\xi) :=
    \log \frac{h(\xi;q^*,q)}{\xi},
\label{Phi(xi)} 
\end{equation}
we have
\begin{equation*}
    \log\frac{h(\xi;q^*,q)}{\xi}
    =
    \frac{1}{2\pi} \int_0^{2\pi}
    \log|q\, h(q^{-1}e^{i\theta};q^*,q)|\,
    \calK_q(\xi,q^{-1}e^{i\theta})\,
    d\theta 
    +ic,
\end{equation*}
and, putting $\xi\mapsto g(z,q)$,
\begin{equation}
    \log\frac{g(z,q^*)}{g(z,q)}
    =
    \frac{1}{2\pi} \int_0^{2\pi}
    \log|q\, h(q^{-1}e^{i\theta};q^*,q)|\,
    \calK_q(g(z,q),q^{-1}e^{i\theta})\,
    d\theta 
    +ic.
\label{log(g/g)}
\end{equation}
It is easy to see that
\[
    1 = 
    \frac{1}{2\pi} \int_0^{2\pi}
    \calK_q(w,q^{-1}e^{i\theta})\,d\theta
\]
for any $w\in\Ann_q$ by a simple residue computation. Multipling
$\log(q^*/q)$ to this equation and adding it to \eqref{log(g/g)}, we
obtain
\begin{equation}
    \log\frac{g(z,q^*)}{g(z,q)} + \log\frac{q^*}{q}
    =
    \frac{1}{2\pi} \int_0^{2\pi}
    \log|q^*\, h(q^{-1}e^{i\theta};q^*,q)|\,
    \calK_q(g(z,q),q^{-1}e^{i\theta})\,
    d\theta 
    +ic.
\label{log(g/g)+log(q/q):temp}
\end{equation}
This implies
\begin{equation}
    c
    =
    - \frac{1}{2\pi} \int_0^{2\pi}
    \log|q^*\, h(q^{-1}e^{i\theta};q^*,q)|\,
    \Im \calK_q(1,q^{-1}e^{i\theta})\,
    d\theta.
\label{c}
\end{equation}
by the normalization $g(1,q)=g(1,q^*)=1$ and
$\log(q^*/q)\in\Real$. Therefore \eqref{log(g/g)+log(q/q):temp} is
rewritten as
\begin{multline}
    \log\frac{g(z,q^*)}{g(z,q)} + \log\frac{q^*}{q}
\\
    =
    \frac{1}{2\pi} \int_0^{2\pi}
    \log|q^*\, h(q^{-1}e^{i\theta};q^*,q)|\,
    \left(
            \calK_q(g(z,q),q^{-1}e^{i\theta})
    - i \Im \calK_q(     1,q^{-1}e^{i\theta})
    \right)
    d\theta,
\label{log(g/g)+log(q/q)}
\end{multline}
which corresponds to \eqref{g(z,s)-g(z,t):chordal} in the chordal case. 

On the other hand, applying \eqref{villat-suppl} to the function
$\Phi(\xi)$, we have
\begin{equation}
    0
    =
    \frac{1}{2\pi}\int_0^{2\pi}
    \log\left|
     \frac{h(q^{-1}e^{i\theta};q^*,q)}{q^{-1}e^{i\theta}}
    \right|\, d\theta
    =
    \log q +
    \frac{1}{2\pi}\int_0^{2\pi}
    \log |h(q^{-1}e^{i\theta};q^*,q)|\, d\theta.
\end{equation}
Adding $\log q^*$ to the above equation, we obtain
\begin{equation}
    \log\frac{q^*}{q}
    =
    \frac{1}{2\pi}\int_0^{2\pi}
    \log |q^* h(q^{-1}e^{i\theta};q^*,q)|\, d\theta,
\label{log(q/q)}
\end{equation}
which substitutes \eqref{t-s:integral-rep} in the chordal case.

Note that $\log|q^* h(q^{-1}e^{i\theta};q^*,q)|=0$ if
$q^{-1}e^{i\theta}$ is mapped to the boundary of $\Ann_{q^*}$ by
$h(\xi;q^*,q)$, namely, if it does not lie on the bold arc of the left
upper annulus in Figure \ref{fig:h(xi;q*,q)}. Since the bold arc shrinks
to $\Lambda(q)$, when $q^*\nearrow q$, the ratio of
\eqref{log(g/g)+log(q/q)} and \eqref{log(q/q)} converges to
\begin{equation}
    \frac{\der \log g(z,q)}{\der \log q} + 1
    =
    \calK_q(g(z,q),\Lambda(q)) - i \Im\calK_q(1,\Lambda(q)),
\label{komatu-loewner:temp}
\end{equation}
where the differentiation in the left hand side is understood as the
left derivative. We can also prove the same differential equation with
the right derivative, and therefore the above equation holds as a usual
differential equation.

The equation \eqref{komatu-loewner:temp} apparently differs from the
Komatu-L\"owner equation \eqref{komatu-loewner}, but we can rewrite
\eqref{komatu-loewner:temp} to \eqref{komatu-loewner}, following the
observation in \cite{Goluzin}. Putting $z=1$ in
\eqref{komatu-loewner:temp}, we have
\[
    1 = \calK_q(1,\Lambda(q)) - i \Im\calK_q(1,\Lambda(q)),
\]
which means $\Re\calK_q(1,\Lambda(q))=1$. Hence, ``$1$'' in the left
hand side of \eqref{komatu-loewner:temp} can be moved to the right hand
side as ``$\Re\calK_q(1,\Lambda(q))$''. Thus we obtain the
Komatu-L\"owner equation,
\begin{equation}
    \frac{\der \log g(z,q)}{\der \log q}
    =
    \calK_q(g(z,q),\Lambda(q)) - \calK_q(1,\Lambda(q)).
\label{komatu-loewner:final}
\end{equation}

\section{One-variable reductions of integrable hierarchies}

\label{section:one}

\subsection{One-variable reductions of the dKP hierarchy}

\subsubsection{The dKP hierarchy}
\label{section:dkphierarchy}

Let ${\bf t}=\{t_1, t_2, t_3, \ldots \}$ be an infinite set of independent variables
(``times'') which are supposed to be real\footnote{In algebraic theory of the dKP
hierarchy (see e.g. \cite{tak-tak:95}) the variables are not necessarily real.}. 
It is convenient to introduce the differential operator
\beq\label{a0}
D(z)=\sum_{k\geq 1} \frac{z^{-k}}{k}\, \p_{t_k}.
\eeq

The dKP hierarchy in the Hirota form can be written as the equation
\beq\label{a1}
e^{D(z_1)D(z_2)F} = 1 - \frac{\p_{t_1}D(z_1)F-\p_{t_1}D(z_2)F}{z_1-z_2}
\eeq
for the real-valued function $F=F({\bf t})$ which should hold for any $z_1$, $z_2$
\cite{take:14-1, tak-tak:95,Teo03}. 
The differential equations of the hierarchy are obtained by expanding this equation
in (inverse) powers of $z_1$, $z_2$. It is convenient to introduce the function
\beq\label{a2}
p(z)=z-\p_{t_1}D(z_1)F,
\eeq
the equation (\ref{a1}) acquires the form
\beq\label{a3}
e^{D(z_1)D(z_2)F} = \frac{p(z_1)-p(z_2)}{z_1-z_2}.
\eeq
The function $p(z)$ clearly depends also on all the times: $p(z)=p(z; {\bf t})$.
One can expand $p(z)$ in a series of the form
\beq\label{a4}
p(z)=z-\frac{u}{z}+\sum_{k\geq 2}\frac{u_k}{z^k},
\eeq
where
\beq\label{a5}
u=u({\bf t})=\p^2_{t_1}F.
\eeq
The coefficients $u_k$ are dependent variables. They are real-valued functions of the times.

If the series $p(z)$ converges in some neighborhood of infinity and defined a function
$p(z)$, it
can be given a geometrical meaning as conformal map from the upper
half plane with a ``fat slit'' to the upper half plane \cite{Z09}.
The times $t_k$ are harmonic moments of the (exterior of) the domain. The coefficient
$u$ in the expansion (\ref{a4}) is called (logarithmic) capacity of the domain. 

Let us take logarithms of both sides of (\ref{a3}),
$$
D(z_1)D(z_2)F=\log \frac{p(z_1)-p(z_2)}{z_1-z_2},
$$
and differentiate with respect to $t_1$. Taking into account (\ref{a2}), we get:
\beq\label{a6}
D(z_2)p(z_1)=-\, \frac{\p_{t_1}p(z_1)-\p_{t_1}p(z_2)}{p(z_1)-p(z_2)},
\eeq
from which it is seen that 
\beq\label{a7}
D(z_2)p(z_1)=D(z_1)p(z_2).
\eeq
Tending $z_2\to \infty$, we get, using (\ref{a4}):
\beq\label{a8}
\p_{t_1}p(z)=-D(z)u.
\eeq

In what follows we assume that the series $p(z)$ converges in some neighborhood of
infinity and defines a meromorphic function (conformal map) $p(z)$. 
Equation (\ref{a6}) acquires a more familiar form in terms of the inverse function to
the $p(z)$, which we denote $z(p)$. It maps the upper half plane to the domain 
which is the complement in the upper half plane to a ``fat slit''. 
Similarly to (\ref{a4}), one can
expand $z(p)$ into a series of the form
\beq\label{a12}
z(p)=p+\frac{u}{p}+\sum_{k\geq 2}v_kp^{-k}.
\eeq

In order to transform equation (\ref{a6})
to a more suggestive form, we introduce the polynomials $B_k(p)$ by the expansion
\beq\label{a13}
-\log \frac{p(z)-p}{z} = \sum_{k\geq 1} \frac{z^{-k}}{k}\, B_k(p).
\eeq
They are called Faber polynomials \cite{Duren}. For example, $B_1(p)=p$. It is easy to see that
\beq\label{a14}
B_k(p)=\Bigl (z^k(p)\Bigr )_{\geq 0},
\eeq
where $(\ldots )_{\geq 0}$ is the polynomial part of the Laurent series 
in $p$ (containing only 
non-negative powers of the variable). Indeed, let $z_1$ be the pre-image of the point 
$p$ under the map $p(z)$, then we can write
$$
\log \Bigl (p(z)-p(z_1)\Bigr )=\log \frac{p(z)-p(z_1)}{z-z_1}+\log (z-z_1)
$$
$$
=\, \log \frac{p(z)-p(z_1)}{z-z_1}+\log z -\sum_{k\geq 1}\frac{z^{-k}}{k}\, z_1^k.
$$
Using the fact that $\displaystyle{\Bigl (\log \frac{p(z)-p(z_1)}{z-z_1}\Bigr )_{\geq 0}=0}$
and comparing with (\ref{a13}), we get (\ref{a14}).

Now, using the relations
\beq\label{a14a}
\p_{t_k}p(z)=-\, \frac{\p_{t_k}z(p)}{\p_p z(p)}, \qquad k\geq 1
\eeq
between the partial derivatives, and, as a consequence,
$$
\p_p z(p) D(z_1)p(z)=-D(z_1) z(p)
$$
(the derivative in the r.h.s. is taken at ``constant $p$'', namely, we differentiate
the coefficients in (\ref{a12})), 
we can rewrite (\ref{a6}) in the form
$$
\p_p z(p) D(z_1)p(z)=-D(z_1) z(p)=\frac{\p_{t_1}z(p) +\p_{t_1}p(z_1) \p_p z(p)}{p(z_1)-p(z)}.
$$
After simple transformations this equation can be written as
\beq\label{a17}
D(z_1)z(p)=\Bigl \{ z(p), \, \log (p(z_1)-p)\Bigr \},
\eeq
where
\beq\label{a17a}
\{f,\, g\}:= \frac{\p f}{\p t_1}\, \frac{\p g}{\p p}- 
\frac{\p g}{\p t_1}\, \frac{\p f}{\p p}
\eeq
is the Poisson bracket. Expanding (\ref{a17}) in powers of $z_1$ and taking into account
(\ref{a13}), we obtain the hierarchy of Lax equations
\beq\label{a18}
\p_{t_k}z(p)=\Bigl \{ B_k(p), \, z(p)\Bigl \}=
\Bigl \{ (z^k(p))_{\geq 0}, \, z(p)\Bigl \}.
\eeq
We see that the conformal map $z(p)$ plays the role of the Lax function. 

\subsubsection{The reduction}

\label{section:dkpreduction}

For one-variable reduction, the dependence of $p(z)$ on the times is implemented by means of 
a single
variable $\lambda = \lambda ({\bf t})$: $p(z; {\bf t})=p(z, \lambda ({\bf t}))$,
i.e., instead of the function of infinitely many independent variables 
$p(z; {\bf t})$ we now deal with a function 
of two variables $p(z, \lambda )$ in which the variable $\lambda$ 
depends on all the times. Our goal is to find a possible form of the functions
$p(z, \lambda )$ such that it would be consistent with the infinite hierarchy. 
The reduction is an exceptional, non-generic solution. We will see that the solutions
correspond to conformal maps from slit domains. 

Using the chain rule of differentiating, we get:
$$
\p_{t_1}p(z)=\p_{\lambda} p (z, \lambda )\cdot \p_{t_1}\lambda , \qquad
D(z_2)p(z_1)=\p_{\lambda }p(z_1, \lambda )\cdot D(z_2)\lambda .
$$
We also have
$$
D(z)\lambda =D(z)u \cdot \frac{\p \lambda}{\p u}=
\frac{D(z)u}{\p_{\lambda }u}=-\frac{\p_{\lambda}p(z, \lambda )}{\p_{\lambda}u}\, 
\p_{t_1}\lambda ,
$$
where we have used equation
(\ref{a8}) at the last step. Substituting all this into (\ref{a6}) and
assuming that $\p_{t_1}\lambda \neq 0$, we get
$$
\p_{\lambda }p(z_1)\p_{\lambda }p(z_2)=
\frac{\p_{\lambda }p(z_1)-\p_{\lambda }p(z_2)}{p(z_1)-p(z_2)}\, \p_{\lambda}u ,
$$
or, after rearranging,
$$
p(z_1)+\frac{\p_{\lambda}u}{\p_{\lambda}p(z_1)} =
p(z_2)+ \frac{\p_{\lambda}u}{\p_{\lambda}p(z_2)}.
$$
It follows from this equation that
\beq\label{a8a}
\xi (\lambda ):=p(z)+\frac{\p_{\lambda}u}{\p_{\lambda}p(z)}
\eeq
does not depend on $z$. This equation can be written as the differential equation
\beq\label{a9}
\p_{\lambda}p(z, \lambda )=-\, \frac{\p_{\lambda}u}{p(z, \lambda )-\xi (\lambda )}
\eeq
for the function $p(z, \lambda )$ with arbitrary function $\xi (\lambda )$. It is called the 
driving function. This function determines the type of the reduction. 
In particular, if one can put 
$\lambda =u$, then the equation simplifies:
\beq\label{a10}
\p_{u}p(z, u )=-\, \frac{1}{p(z, u )-\xi (u )}.
\eeq
All the coefficients $u_k$ in (\ref{a4}) become functions of $u$. 

Equation (\ref{a9}) is the (chordal) L\"owner equation for 
a one-parameter family of conformal maps $p(z, \lambda )$ 
from the upper half plane with a curved slit starting at 
the real axis to the upper half plane. Here $z$ is the coordinate in the 
upper half plane with a slit and $\lambda$ is the parameter of the curved slit.
As we have seen in section \ref{section:chordal}, 
the form of the slit is determined by the function 
$\xi (\lambda)$, the variable $\lambda$ being a parameter along the curve (for example, the length
of the curve). One can say, therefore, that the type of the reduction is determined by
the form of the slit. 
The choice $\lambda=u$ corresponds
to the choice of capacity as a natural parameter along the curved slit. 

We see that one-variable reductions of the dKP hierarchy are determined 
by the form of the slit if it comes from such a family
of conformal maps and are given by solutions
of the chordal L\"owner equation (\ref{a9}). 

The differential equation (\ref{a9}) can be equivalently written as 
a partial differential equation for the Lax function $z(p)$ which maps 
the upper half plane to the upper half plane with a slit. Indeed, since
$$
\p_{\lambda}p(z)= -\, \frac{\p_{\lambda}z(p)}{\p_p z(p)},
$$
we represent equation (\ref{a9}) in the equivalent form
\beq\label{a11}
\frac{\p z(p)}{\p \lambda}=\frac{\p_{\lambda}u}{p-\xi(\lambda)}\,
\frac{\p z(p)}{\p p}
\eeq
which is also referred to as chordal L\"owner equation. We see that the Lax function
satisfies the chordal L\"owner equation of this type.

Given a one-variable reduction, it is not difficult to obtain the solution to the hierarchy.
Using (\ref{a8}), we have:
$$
D(z)\lambda =D(z)u \cdot \p_u\lambda =-\p_{t_1}p(z)\p_u\lambda =
-\p_{\lambda}p(z)\, \p_{t_1}\lambda \, \p_u\lambda
=\, \frac{\p_{t_1}\lambda}{p(z)-\xi(\lambda)},
$$
where we have used the L\"owner equation (\ref{a9}) at the last step. Differentiating 
(\ref{a13}) with respect to $p$, we can write
$$
D(z)\lambda =\sum_{k\geq 1}\frac{z^{-k}}{k}\, B_k'(\xi(\lambda)) \p_{t_1}\lambda,
$$
where $B_k'(p)=\p_p B_k(p)$ is the derivative of the Faber polynomial, or
\beq\label{a15}
\p_{t_k}\lambda = B_k'(\xi(\lambda))\, \p_{t_1}\lambda , \quad k\geq 1,
\eeq
which is a one-component differential equation of the hydrodynamic type.
The implicit solution for $\lambda =\lambda ({\bf t})$ 
of the hodograph relation
\beq\label{a16}
t_1+\sum_{k\geq 2}t_k B_k'(\xi(\lambda))=R(\lambda)
\eeq
gives a solution of (\ref{a15}). Here $R(\lambda)$ is an arbitrary function.  

\subsection{One-variable reductions of the dBKP hierarchy}

\subsubsection{The dBKP hierarchy}

For the theory of the dBKP hierarchy see \cite{BK05,CT06,T06}.
Let ${\bf t}=\{t_1, t_3, t_5, \ldots \}$ be an infinite set of independent variables
(``times'') indexed by odd natural numbers. Here we suppose that they are real. It is convenient
to introduce the differential operator
\beq\label{bkp1}
D^{\rm o}(z)=\sum_{k\geq 1, \, {\rm odd}}\! \frac{z^{-k}}{k}\, \p_{t_k}.
\eeq
The dBKP hierarchy in the Hirota form can be written as the equation
\beq\label{bkp2}
1-2\, \frac{D^{\rm o}(z_1)\p_{t_1}F -D^{\rm o}(z_2)\p_{t_1}F}{z_1-z_2}=
\left (1-2\, \frac{D^{\rm o}(z_1)\p_{t_1}F +D^{\rm o}(z_2)\p_{t_1}F}{z_1+z_2}\right )
e^{4D^{\rm o}(z_1)D^{\rm o}(z_2)F}
\eeq
for the real-valued function $F=F({\bf t})$ which should be valid for any $z_1, z_2$. 
In terms of the (odd) function
\beq\label{bkp3}
p(z)=z-2D^{\rm o}(z)\p_{t_1}F, \quad p(-z)=-p(z)
\eeq
the equation (\ref{bkp2}) reads
\beq\label{bkp4}
\frac{p(z_1)-p(z_2)}{z_1-z_2}=\frac{p(z_1)+p(z_2)}{z_1+z_2}\,
e^{4D^{\rm o}(z_1)D^{\rm o}(z_2)F}.
\eeq
The expansion of the function $p(z)$ in the Laurent series is
\beq\label{bkp5}
p(z)=z-\frac{u}{z}+\sum_{k\geq 3, \, {\rm odd}}\frac{u_k}{z^k},
\eeq
where
\beq\label{bkp6}
u=u({\bf t}) =2\p_{t_1}^2F.
\eeq

Taking logarithm of equation (\ref{bkp4}), differentiating with respect to $t_1$ and using
the definition (\ref{bkp3}), we obtain the equation
\beq\label{bkp7}
2D^{\rm o}(z_1)p(z_2)=\p_{t_1} \log \frac{p(z_1)+p(z_2)}{p(z_1)-p(z_2)}.
\eeq
from which it follows that $D^{\rm o}(z_1)p(z_2)=D^{\rm o}(z_2)p(z_1)$ (this follows also
from the definition (\ref{bkp3})). Tending $z_2 \to \infty$, we get
\beq\label{bkp8}
\p_{t_1}p(z)=-D^{\rm o}(z)u.
\eeq

Let us rewrite equation (\ref{bkp7}) in terms of the function $z(p)$, inverse to the $p(z)$
(like $p(z)$, it is an odd function with the Laurent series of the form 
$z(p)=p+ O(p^{-1})$).
The calculation is similar to the one leading to equation (\ref{a17}). 
Using the relation (\ref{a14a}), we get, after simple transformations:
\beq\label{bkp9}
2D^{\rm o}(z_1)z(p)=\left \{ z(p), \, \log \frac{p(z_1)-p}{p(z_1)+p}\right \},
\eeq
where $\{\, , \, \}$ is the Poisson bracket (\ref{a17a}). This is the generating Lax equation
for the dBKP hierarchy, $z(p)$ being the Lax function. Expanding equation (\ref{bkp9}) in 
powers of $z_1$, one obtains the hierarchy of Lax equations through the Faber polynomials. 
The Faber polynomials $B_k(p)$ 
are introduced by the same formulas (\ref{a13}), (\ref{a14}) as 
in section \ref{section:dkphierarchy}. The fact that $p(z)$ is an odd function implies that
$B_k(-p)=(-1)^k B_k(p)$ and we have the expansion
\beq\label{bkp9a}
\log \frac{p(z)+p}{p(z)-p}=2\! \sum_{k\geq 1, \, {\rm odd}} \frac{z^{-k}}{k}\, B_k(p), 
\quad B_k(p)=\Bigl (z^k(p)\Bigr )_{\geq 0}.
\eeq
The Lax equations are of the form (\ref{a18}). 

We note that the dispersionless limit of the CKP hierarchy is known to be the same as for the 
BKP hierarchy (but the dispersionful hierarhies are different), 
so it is given by the same equation (\ref{bkp2}). 

\subsubsection{The reduction}

The definition of the one-variable reduction is the same as in section \ref{section:dkpreduction}.
Substituting the reduction condition into equation (\ref{bkp7}), we get:
$$
\p_{\lambda}p(z_1)D^{\rm o}(z_2)\lambda =
\frac{p(z_1)\p_{\lambda}p(z_2)-p(z_2)\p_{\lambda}p(z_1)}{p^2(z_1)-p^2(z_2)}\, \p_{t_1}\lambda.
$$
Next, plugging here the relation 
$$
D^{\rm o}(z_2)\lambda =-\frac{\p_{\lambda}p(z_2)}{\p_{\lambda}u}\, \p_{t_1}\lambda
$$
(a consequence of (\ref{bkp8}) and the reduction condition) and assuming that 
$\p_{t_1}\lambda$ is not identically zero, we obtain:
$$
\p_{\lambda}p(z_1)\p_{\lambda}p(z_2)=
\frac{p(z_2)\p_{\lambda}p(z_1)-p(z_1)\p_{\lambda}p(z_2)}{p^2(z_1)-p^2(z_2)}\, \p_{\lambda}u
$$
or, after, rearranging,
$$
p^2(z_1)+\frac{p(z_1)\p_{\lambda}u}{\p_{\lambda}p(z_1)}=
p^2(z_2)+\frac{p(z_2)\p_{\lambda}u}{\p_{\lambda}p(z_2)}.
$$
It follows from this relation that
\beq\label{bkp10}
\xi^2(\lambda):= p^2(z)+\frac{p(z)\p_{\lambda}u}{\p_{\lambda}p(z)}
\eeq
does not depend on $z$. This is equivalent to the differential equation
\beq\label{bkp11}
\p_{\lambda}p(z)=-\frac{p(z)\, \p_{\lambda}u}{p^2(z)-\xi^2(\lambda )}
\eeq
or
\beq\label{bkp11a}
\p_{\lambda}p(z)=-\frac{\p_{\lambda}u/2}{p(z)-\xi(\lambda)}-
\frac{\p_{\lambda}u/2}{p(z)+\xi(\lambda)}.
\eeq
The corresponding equation for the inverse function is
\beq\label{bkp12}
\p_{\lambda}z(p)=\frac{p\, \p_{\lambda} u}{p^2 -\xi^2(\lambda)}\, \p_p z(p).
\eeq
This is the quadrant L\"owner equation \cite{T13} with the driving function $\xi(\lambda)$. 
Suitably normalized families of conformal maps of slit domains in the quadrant are
solutions to this equation. 

Using the quadrant L\"owner equation, we can write
$$
D^{\rm o}(z)\lambda =\frac{1}{2}\left (\frac{1}{p(z)-\xi(\lambda)}+
\frac{1}{p(z)+\xi(\lambda)}\right )\p_{t_1}\lambda .
$$
Using the expansion (\ref{bkp9a}), we obtain from here the system of partial differential
equations for $\lambda$ of the hydrodynamic type:
\beq\label{bkp13}
\p_{t_k}\lambda =B_k'(\xi(\lambda))\p_{t_1}\lambda , \quad k=1,3,5, \ldots \, 
\eeq
with the general solution in the hodograph form
\beq\label{bkp14}
t_1+\! \sum_{k\geq 3, \, {\rm odd}}B_k'(\xi(\lambda))=R(\lambda),
\eeq
where $R(\lambda)$ is an arbitrary function. 

\subsection{One-variable reductions of the dToda hierarchy}

\subsubsection{The dToda hierarchy}

For the dToda hierarchy, there are 
two sets of infinitely many independent variables (``times''): complex variables 
${\bf t}=\{t_1, t_2, t_3, \ldots \}$ and
$\bar {\bf t}=\{\bar t_1, \bar t_2, \bar t_3, \ldots \}$.
Here we take $\bar {\bf t}$ not independent from ${\bf t}$ but assume that
they are complex conjugate
to ${\bf t}$. There is also a real variable $t_0$.
Similarly to (\ref{a0}), we introduce the differential operators
$$
D(z)=\sum_{k\geq 1} \frac{z^{-k}}{k}\, \p_{t_k}, \quad
\bar D(\bar z)=\sum_{k\geq 1} \frac{\bar z^{-k}}{k}\, \p_{\bar t_k}.
$$

The dToda hierarchy can be written as the following system of equations:
\beq\label{b1}
\left \{
\begin{array}{l}
(z_1-z_2)e^{D(z_1)D(z_2)F} =z_1e^{-\p_{t_0}D(z_1)F}-z_2e^{-\p_{t_0}D(z_2)F},
\\ \\
z_1\bar z_2 \Bigl (1-e^{-D(z_1)\bar D(\bar z_2)F}\Bigr )=
e^{\p_{t_0}(\p_{t_0}+D(z_1)+\bar D(\bar z_2))F}
\end{array}\right.
\eeq
for a real-valued function $F=F({\bf t}, t_0, \bar {\bf t})$, 
which should hold for any $z_1$, $z_2$. Besides, there is the equation which is
complex-conjugate to the first equation in (\ref{b1}). 
It is convenient to introduce the function
\beq\label{b2}
w(z)=z\exp \Bigl (-\frac{1}{2}\, \p_{t_0}^2 F-\p_{t_0}D(z)F\Bigr ),
\eeq
then the equations of the dToda hierarchy can be written in the form
\beq\label{b3}
\left \{
\begin{array}{l}
\displaystyle{
\frac{w(z_1)-w(z_2)}{z_1-z_2} =e^{-\frac{1}{2}\, \p_{t_0}^2F +D(z_1)D(z_2)F}},
\\ \\
\displaystyle{1-\Bigl (w(z_1)\bar w(\bar z_2)\Bigr )^{-1}=e^{-D(z_1)\bar D(\bar z_2)F}}
\end{array}\right.
\eeq
In the second equation $\bar w(\bar z)=\overline{w(z)}$. 
The function $w(z)$ depends also on
all the times: $w(z)=w(z; {\bf t}, t_0, \bar {\bf t})$.
The expansion of the function $w(z)$ into the series has the form
\beq\label{b4}
w(z)=\frac{z}{r} +\sum_{k\geq 0}\frac{u_k}{z^k},
\eeq
where $r$ is a real quantity given by
\beq\label{b5}
\log r=\frac{1}{2}\, \p^2_{t_0}F.
\eeq
Note that
\beq\label{c1}
D(z)\log r = \frac{1}{2}\, \p_{t_0}^2 D(z)F=-\frac{1}{2}\, \p_{t_0}\log (rw(z)).
\eeq
Another way to derive equation (\ref{c1}) is as follows.
Consider the first equation in (\ref{b3}). 
Let us write it in the form
\beq\label{b3a}
e^{(\p_{t_0}+D(z_1))(\p_{t_0}+D(z_2))F}=\frac{rw^{-1}(z_1)-rw^{-1}(z_2)}{z_1^{-1}-z_2^{-1}}.
\eeq
Note that this equation already contains the definition of the function $w(z)$ (\ref{b2})
as the limit case $z_2\to \infty$. 
Let us take logarithm of the both sides of (\ref{b3a}) and differentiate
with respect to $t_0$. Taking into account (\ref{b2}), we get:
\beq\label{b6}
(\p_{t_0}+D(z_2))\log \frac{w(z_1)}{rz_1}=-\, 
\frac{\p_{t_0}(rw^{-1}(z_1))-\p_{t_0}(rw^{-1}(z_2))}{r(w^{-1}(z_1)-w^{-1}(z_2))}.
\eeq
The right hand side is symmetric with respect to permutation of $z_1, z_2$, whence
\beq\label{b7}
(\p_{t_0}+D(z_2))\log (rw^{-1}(z_1))=(\p_{t_0}+D(z_1))\log (rw^{-1}(z_2)).
\eeq
Tending $z_2\to \infty$, we get, using (\ref{b4}), that
\beq\label{b8}
\p_{t_0}\log (rw^{-1}(z))=(\p_{t_0}+D(z))\log r^2
\eeq
which is the same as (\ref{c1}). 

In what follows we assume that the series (\ref{b4}) converges in a neighborhood
of infinity and defines there a meromorphic function. This function 
$w(z)$ has the meaning of the conformal map of a domain containing 
$\infty$ to the exterior of the unit circle normalized in such a way that 
$w(\infty)=\infty$ and 
$\mbox{arg}\, w'(\infty)>0$. The times $t_k, \bar t_k$ 
are harmonic moments of the (exterior of) the domain 
and $t_0$ is proportional to its area \cite{WZ00,Z01}. 
The quantity $r>0$ is called the conformal radius of the domain.

Taking logarithm of equations (\ref{b3}), differentiating with respect to $t_0$ and
using (\ref{c1}), we obtain the equations
\beq\label{c2}
\begin{array}{l}
D(z_1)\log w(z)=-\p_{t_0}\log \Bigl (w(z_1)-w(z)\Bigr ) +\frac{1}{2}\, \p_{t_0}\log \Bigl (
z_1w(z_1)/r\Bigr ),
\\ \\
\bar D(\bar z_1)\log w(z)=\p_{t_0}\log \Bigl (\bar w(\bar z_1)-w^{-1}(z)\Bigr ) -
\frac{1}{2}\, \p_{t_0}\log \Bigl (
\bar z_1\bar w(\bar z_1)/r\Bigr ).
\end{array}
\eeq
It is instructive to rewrite them in terms of the inverse function to $w(z)$ which we
denote as $z(w)$ and the analytic continuation $\bar z(w^{-1})$ of the complex conjugate
function $\overline{z(w)}$ from the unit circle. They have the expansions
\beq\label{c2a}
z(w)=rw +\sum_{k\geq 0}v_k w^{-k}, \quad \bar z(w^{-1})=rw^{-1}+\sum_{k\geq 0}
\bar v_k w^k.
\eeq
Noting the relations
$$
\p_{t_k}w(z)=-\, \frac{\p_{t_k}z(w)}{\p_w z(w)}, \quad k\geq 0
$$
between the partial derivatives, and, as a consequence,
$$
w\p_w z(w) D(z_1)\log w(z)=-D(z_1) z(w)
$$
(the derivative in the r.h.s. is taken at constant $w$), 
we can rewrite the first equation in (\ref{c2}) in the form
$$
\frac{D(z_1)z(w)}{w\p_w z(w)}=\frac{\p_{t_0}w(z_1)+\p_{t_0}z(w)/\p_wz(w)}{w(z_1)-w(z)}-
\frac{1}{2}\, \p_{t_0}\, \log \Bigl (z_1w(z_1)/r\Bigr )
$$
or
\beq\label{c3}
D(z_1)z(w)=\left \{ z(w), \, \log \Bigl (w(z_1)-w\Bigr )+\frac{1}{2}\, \log \frac{r}{z_1w(z_1)}
\right \}_{\rm Toda},
\eeq
where
\beq\label{c4}
\{f, g\}_{\rm Toda}=w\frac{\p f}{\p w}\frac{\p g}{\p t_0}-w
\frac{\p f}{\p t_0}\frac{\p g}{\p w}
\eeq
is the Poisson bracket for the dToda hierarchy. Similarly, the second equation in (\ref{c2})
can be written in the form
\beq\label{c5}
\bar D(\bar z_1)z(w)=-\left \{ z(w), \, \log \Bigl (\bar w(\bar z_1)-w^{-1}
\Bigr )+\frac{1}{2}\, \log \frac{r}{\bar z_1\bar w(\bar z_1)}
\right \}_{\rm Toda}.
\eeq
Equations (\ref{c3}), (\ref{c5}) are generating Lax equations for the dToda hierarchy.
To see this, let us introduce the Faber polynomials $A_k(w)$, $\bar A_k(w^{-1})$
according to the expansions
\beq\label{c6}
\begin{array}{l}
\displaystyle{-\log \Bigl (w(z)-w\Bigr )+\frac{1}{2}\, \log \frac{zw(z)}{r}=
\sum_{k\geq 1}\frac{z^{-k}}{k}\, A_k(w),}
\\ \\
\displaystyle{-\log \Bigl (\bar w(\bar z)-w^{-1}\Bigr )+\frac{1}{2}\, 
\log \frac{\bar z\bar w(z)}{r}=
\sum_{k\geq 1}\frac{\bar z^{-k}}{k}\, \bar A_k(w^{-1}),}
\end{array}
\eeq
then equations (\ref{c3}), (\ref{c5}) imply the Lax equations of the dToda hierarchy:
\beq\label{c8}
\p_{t_k}z(w)=\Bigl \{ A_k(w), \, z(w)\Bigr \}_{\rm Toda}, \quad
\p_{\bar t_k}z(w)=-\Bigl \{ \bar A_k(w^{-1}), \, z(w)\Bigr \}_{\rm Toda}.
\eeq
It is easy to see that
\beq\label{c7}
A_k(w)=\Bigl (z^k(w)\Bigr )_{>0}+\frac{1}{2}\Bigl (z^k(w)\Bigr )_{0}, \quad
\bar A_k(w^{-1})=\Bigl (\bar z^k(w^{-1})\Bigr )_{<0}+\frac{1}{2}\Bigl (\bar z^k(w^{-1})\Bigr )_{0},
\eeq
where $(\ldots )_{>0}$, $(\ldots )_{<0}$, $(\ldots )_{0}$ mean respectively 
positive, negative and constant parts of a Laurent series. To see this, we fix a point
$w_1=z(w_1)$ and write
$$
\log \left (1-\frac{z(w)}{z(w_1)}\right )=
\log \left (1-\frac{w}{w(z_1)}\right ) +\log \frac{rw(z_1)}{z_1}+\log
\frac{z(w_1)-z(w)}{r(w_1-w)}.
$$
Now, we notice that the expansion of the first (third) term contains only positive
(negative) powers of $w$ while the rest is just the constant term. Equations (\ref{c7}) 
easily follow from this.



\subsubsection{The reduction}

\label{section:dtodareduction}

The one-variable reduction is defined similarly to the case of the
dKP hierarchy.
For the one-variable reduction, the dependence of $w(z)$ on all the times is implemented 
by a single function $\lambda = \lambda ({\bf t}, t_0, \bar {\bf t})$: 
$w(z; {\bf t}, t_0, \bar {\bf t})=w(z, \lambda ({\bf t}, t_0, \bar {\bf t}))$,
i.e., instead of a function of infinitely many variables $w(z; {\bf t}, t_0, \bar {\bf t})$ 
we deal with a function of two variables 
$w(z, \lambda )$, where the variable $\lambda$ 
is a function of all the times. 
Here we again assume that the formal series $w(z, \lambda)$ gives a one-parameter family of
functions in some domain in the exterior of the unit disk. 
Our goal is 
to find a possible form of the functions
$w(z, \lambda )$ such that it would be consistent with the dToda hierarchy.

In the calculations below we closely follow \cite{TTZ06,Z07}.
Using the chain rule of differentiating, we have:
$$
\p_{t_0}(rw^{-1}(z))=\p_{\lambda} (rw ^{-1}(z, \lambda ))\cdot \p_{t_0}\lambda , 
$$
$$
(\p_{t_0}+D(z_2))\log (rw^{-1}(z_1))=\p_{\lambda }\log (rw^{-1}(z_1, \lambda ))
\cdot (\p_{t_0}+D(z_2))\lambda 
$$
and
$$
(\p_{t_0}+D(z))\lambda =(\p_{t_0}+D(z))\log r \cdot \frac{\p \lambda}{\p \log r}
$$
$$
=\frac{(\p_{t_0}+D(z))\log r}{\p_{\lambda }\log r}=\frac{\p_{\lambda}
\log (rw^{-1}(z, \lambda ))}{2\p_{\lambda}\log r}\, 
\p_{t_0}\lambda ,
$$
where we have used (\ref{b8}) in the last step. Substituting this into (\ref{b6})
and assuming that $\p_{t_0}\lambda \neq 0$, we get:
$$
\p_{\lambda }\log (rw^{-1}(z_1))\p_{\lambda }\log (rw^{-1}(z_2))=
2\,
\frac{\p_{\lambda }(rw^{-1}(z_1))-\p_{\lambda }(rw^{-1}(z_2))}{rw^{-1}(z_1)-
rw^{-1}(z_2)}\, \p_{\lambda}\log r \,
$$
or, after rearranging, 
$$
-w(z_1)+\frac{2w(z_1)\p_{\lambda}\log r}{\p_{\lambda}\log (rw^{-1}(z_1))}=
-w(z_2)+\frac{2w(z_2)\p_{\lambda}\log r}{\p_{\lambda}\log (rw^{-1}(z_2))}.
$$
It follows from this equation that
\beq\label{varphi}
\eta (\lambda ):=w(z)-\frac{2w(z)\p_{\lambda}\log r}{\p_{\lambda}\log (rw^{-1}(z))}=
-w(z)\, \frac{\p_{\lambda}\log (rw(z))}{\p_{\lambda}\log (rw^{-1}(z))}
\eeq
does not depend on $z$. This relation can be rewritten as the differential equation
\beq\label{b9}
\p_{\lambda}\log w(z, \lambda )=-\, \frac{w(z, \lambda)+
\varphi(\lambda)}{w(z, \lambda )-\eta (\lambda )}\, \p_{\lambda}\log r
\eeq
for the function $w(z, \lambda )$ with an arbitrary function $\eta (\lambda )$.
In particular, one can put
$\lambda =\log r$, then the equation simplifies:
\beq\label{b10}
\frac{\p \log w(z)}{\p \log r}=-\, \frac{w(z)+\eta (r) }{w(z)-\eta (r)}
\eeq
and all the coefficients $u_k$ in (\ref{b4}) become functions of $r$. 

Let us now see what the second equation in (\ref{b3}) gives. We write it in the form
\beq\label{b11}
\log \Bigl (w(z_1)\overline{w(z_2)}-1\Bigr )=
z_1\bar z_2 -(\p_{t_0}+D(z_1))(\p_{t_0}+\bar D(\bar z_2))F
\eeq
and differentiate both sides with respect to $t_0$. We obtain:
$$
\frac{\p_{t_0}w(z_1)\, \overline{w(z_2)}+\p_{t_0}
\overline{w(z_2)}\, w(z_1)}{w(z_1)\overline{w(z_2)}-1}
=-(\p_{t_0}+D(z_1))\log \Bigl (r\overline{w^{-1}(z_2)}\Bigr ).
$$
Substituting here the reduction condition and using the relations obtained above,
we get:
$$
2\p_{\lambda}\log r \,
\frac{\p_{\lambda}w(z_1)\, 
\overline{w(z_2)}+\p_{\lambda}\overline{w(z_2)}\, w(z_1)}{w(z_1)\overline{w(z_2)}-1}=
\p_{\lambda}\log (rw^{-1}(z_1))\p_{\lambda}\log (r\overline{w^{-1}(z_2)}).
$$
After simple transformations this equation can be represented in the form
\beq\label{b12}
\left (w(z_1)-\frac{2w(z_1)\p_{\lambda}\log r}{\p_{\lambda}\log (rw^{-1}(z_1)}\right )
\left (\overline{w(z_2)}-\frac{2\overline{w(z_2)}
\p_{\lambda}\log r}{\p_{\lambda}\log (r\overline{w^{-1}(z_2)}}\right )=1.
\eeq
The definition (\ref{varphi}) implies that this equation means that
\beq\label{b13}
\eta (\lambda )\overline{\eta(\lambda)}=|\eta(\lambda)|^2=1,
\eeq
i.e., $\eta(\lambda)$ has the form
\beq\label{b14}
\eta(\lambda)=e^{i\xi(\lambda)}
\eeq
with a real-valued function $\xi(\lambda)$. 

Therefore, our differential equation acquires the form
\beq\label{b15}
\p_{\lambda}w(z, \lambda )=-w(z, \lambda ) \, \frac{w(z, \lambda)+
e^{i\xi (\lambda)}}{w(z, \lambda )-e^{i\xi (\lambda )}}\, \p_{\lambda}\log r.
\eeq
The equation (\ref{b15}) is the (radial) L\"owner 
equation (\ref{le3}) for the conformal map $w(z)$ from the exterior of the unit disk
with a slit of arbitrary form to the exterior of the unit disk. 
The form of the slit is parametrized by the driving function $\xi (\lambda)$ which
simultaneously determines the type of the reduction. 
Note that in (\ref{le3}) we assumed that $r=e^t$. The last factor $\p_{\lambda}\log r$ 
comes from a different parametrization. 

Introducing $p(z, \lambda)=-i\log w(z, \lambda)$, one can write the radial L\"owner
equation in the form
\beq\label{b16}
\p_{\lambda}p(z, \lambda)= \cot \Bigl (\frac{p(z, \lambda )-\xi(\lambda)}{2}\Bigr )\, 
\p_{\lambda}\log r
\eeq
which is the trigonometric extension of the chordal (rational) 
L\"owner equation (\ref{a9}). The function $p(z, \lambda)$ maps a half-strip
with a slit onto the half-strip. 

Note also that equation (\ref{b15}) can be equivalently rewritten as the 
partial differential equation
\beq\label{b17}
\p_{\lambda}z(w)=\frac{w+e^{i\xi(\lambda)}}{w-e^{i\xi(\lambda)}}\, \p_{\lambda}\log r \,
w\p_w z(w)
\eeq
for the Lax function $z(w)$ (the inverse of $w(z)$).

The dependence of $\lambda$ on the times $t_k$, $\bar t_k$ is determined by a system of
equations of the hydrodynamic type. They follow from equation (\ref{c1}) which can be written as
$$
D(z)\lambda =\frac{D(z)\log r}{\p_{\lambda}\log r}=-\frac{1}{2}
\left (1+\frac{\p_{\lambda}\log w(z)}{\p_{\lambda}\log r}\right ) \p_{t_0}\lambda ,
$$
taking into account the reduction condition. Using the L\"owner equation (\ref{b15}), we obtain:
$$
D(z)\lambda =\frac{\eta(\lambda)\p_{t_0}\lambda}{w(z)-\eta(\lambda)}.
$$
From (\ref{c6}) we conclude that
\beq\label{b17a}
\frac{\eta}{w(z)-\eta}=\sum_{k\geq 1}\frac{z^{-k}}{k}\, \phi_k(\eta),
\qquad \phi_k(w)=w\p_w A_k(w).
\eeq
Therefore, the system of hydrodynamic type reads
\beq\label{b18}
\p_{t_k}\lambda =\phi_k(\eta (\lambda))\p_{t_0}\lambda .
\eeq
Equations containing $\bar t_k$-derivatives are obtained by complex conjugation
of (\ref{b18}). The hodograph equation
\beq\label{b19}
t_0 + \sum_{k\geq 1}t_k \phi_k(\eta (\lambda ))+
\sum_{k\geq 1}\bar t_k \overline{\phi_k(\eta (\lambda ))}=R(\lambda),
\eeq
gives solutions in an implicit form. Here
$R(\lambda )$ is an arbitrary function.

\subsection{One-variable reductions of the dDKP hierarchy}
\label{subsec:1-var-reduction:dDKP}

The DKP hierarchy (known also as Pfaff-KP) 
is an integrable hierarchy with 
$D_{\infty}$ symmetry. It was first introduced by 
M.Jimbo and T.Miwa in 1983 \cite{JimboMiwa} and it
was subsequently rediscovered under the names
coupled KP hierarchy \cite{HO} and
the Pfaff lattice \cite{AHM,ASM}, see also
\cite{Kakei,IWS,Willox}. 
The solutions and the algebraic structure were studied in 
\cite{Kodama,Kodama1,AKM}, the relation to matrix integrals 
was elaborated in \cite{AHM,ASM,Kakei,Vandeleur,Orlov}.

\subsubsection{Algebraic form of the dDKP  hierarchy}

The dispersionless version of the DKP hierarchy (the dDKP
hierarchy) was suggested in 
\cite{Takasaki07,Takasaki09}.
It is an infinite system
of differential equations for a function 
$F=F({\bf t})$ of the infinite number of 
``times''
${\bf t}=\{t_0, t_1, t_2, \ldots \}$. Here again we assume that
$F$ is a real-valued function of real variables ${\bf t}$. 
The differential equations are obtained by expanding 
equations
\beq\label{D1}
e^{D(z_1)D(z_2 )F}\left (1-\frac{1}{z_{1}^2z_{2}^2}\,
e^{2\p_{t_0}(2\p_{t_0} + D(z_1) + D(z_2 ))F}\right )=
1-\frac{\p_{t_1}D(z_1)F -\p_{t_1}D(z_2 )F}{z_1-z_2},
\eeq
\beq\label{D2}
e^{-D(z_1)D(z_2 )F}
\, \frac{z_{1}^2 e^{-2\p_{t_0}D(z_1)F}-z_{2}^2 e^{-2\p_{t_0}D(z_2 )F}}{z_1-z_2}
=z_1 +z_2 -\p_{t_1}\! \Bigl (2\p_{t_0} +D(z_1)+D(z_2 )\Bigr )F,
\eeq
where $D(z)$ is the operator (\ref{a0}),
in powers of $z_1$, $z_2$.

In this section we use the differential operator
\beq\label{E5}
\nabla (z)=\p_{t_0}+ D(z)
\eeq
which in the dDKP case is more convenient than $D(z)$. 
Introducing the functions
\beq\label{D4}
p(z)=z-\p_{t_1} \nabla (z)F,
\qquad
w(z)=z^2 e^{-2\p_{t_0}\nabla (z)F},
\eeq
we can rewrite equations (\ref{D1}), (\ref{D2}) in a more compact form
\beq\label{D1a}
e^{D(z_1)D(z_2 )F}\left (1-\frac{1}{w(z_1)w(z_2 )}\right )=
\frac{p(z_1)-p(z_2 )}{z_1-z_2}\,,
\eeq
\beq\label{D2a}
e^{-D(z_1)D(z_2 )F + 2\p_{t_0}^2F}
\, \, \frac{w(z_1)-w(z_2)}{z_1-z_2}=p(z_1)+p(z_2).
\eeq
Multiplying the two equations, we observe that 
$e^{D(z_1)D(z_2 )F}$ cancels and we get the relation
$$
p^2(z_1)-e^{2F_{00}}\Bigl (w(z_1)+w^{-1}(z_1)\Bigr )=
p^2(z_2 )-e^{2F_{00}}\Bigl (w(z_2 )+w^{-1}(z_2 )\Bigr )
$$
from which it follows that $p^2(z)-e^{2F_{00}}\Bigl (w(z)+w^{-1}(z)\Bigr )$
does not depend on $z$ 
(here and below we use the short-hand notation $F_{mn}=\p_{t_m} \p_{t_n}F$). 
Tending $z$ to infinity, we find that this 
expression is equal to 
$F_{02} -2F_{11} -F_{01}^2$.
Therefore, we conclude that the functions $p(z), w(z)$ satisfy the
algebraic equation \cite{Takasaki09}
\beq\label{D5}
p^2(z)=R^2 \Bigl (w(z)+w^{-1}(z)\Bigr )+V\,,
\eeq
where
\beq\label{D6}
R=e^{F_{00}}, \qquad 
V=F_{02} -2F_{11} -F^2_{01}
\eeq
are real numbers depending on the times ($R$ is positive).
The equation (\ref{D5}) defines a family of elliptic curves, parametrized by 
$R$ and $V$, with
$p$, $w$ being algebraic functions on the curve.

\subsubsection{Elliptic form of the dDKP hierarchy}

Remarkably, the dDKP hierarchy admits an elliptic reformulation 
suggested in \cite{akh-zab:14-1}, see also 
\cite{akh-zab:14-2,ATZ17}.
The first step in this reformulation is uniformization of the curve (\ref{D5}) through  
elliptic functions. 
To this end, we use the standard Jacobi
theta functions $\theta_a (u)=\theta_a (u, \tau )$ ($a=1,2,3,4$).
Their definition is given in Appendix A. 

The uniformization of the relation (\ref{D5}) is as follows:
\beq\label{E1}
p(z)=\gamma \, \theta_4^2(0)\, \frac{\theta_2(u(z))\,
\theta_3(u(z))}{\theta_1(u(z))\, \theta_4(u(z))}\,,
\qquad
w(z)=\left (\frac{\theta_4(u(z))}{\theta_1(u(z))}\right )^2\,, 
\eeq
where $u(z)=u(z, {\bf t})$ is some function 
of $z$, $\gamma$ is a $z$-independent
factor, and
\beq\label{E2}
R=\gamma\, \theta_2(0)\, \theta_3(0)\,, \qquad
V=-\gamma^2 \Bigl ( \theta_2^4(0)+\theta_3^4(0)\Bigr ).
\eeq
In this parametrization, the equation of the curve (\ref{D5}) is equivalent to
the identity
$$
\theta_4^2(0)\, \frac{\theta_2^2(u)\,
\theta_3^2(u)}{\theta_1^2(u)\, \theta_4^2(u)}=
\theta_2^2(0)\theta_3^2(0)\! \left (\frac{\theta_4^2(u)}{\theta_1^2(u)}+
\frac{\theta_1^2(u)}{\theta_4^2(u)}\right )-
\Bigl (\theta_2^4(0)+\theta_3^4(0)\Bigr )
$$
which can be proved either by using some standard identities
for theta-functions or by comparing analytical properties 
of the both sides.

Note that $\gamma$ as well as the modular parameter $\tau$ depend on the times:
$\gamma =\gamma ({\bf t})$, $\tau =\tau ({\bf t})$.
The reality 
of the coefficients  $R^2$, $V$ 
implies certain restrictions on possible 
values of $\tau$.
In what follows we assume that $\tau$ is purely imaginary.
We normalize the function $u(z)$ by the condition 
$u(\infty )=0$, with the expansion around $\infty$ being
\beq\label{E3}
u(z, {\bf t})=\frac{c_1({\bf t})}{z}+\sum_{k\geq 2}\frac{c_k({\bf t})}{z^k} 
\eeq
with real coefficients $c_k$.

It is not difficult to check the identity
$$
\frac{w(z_1)-w(z_2)}{p(z_1)+p(z_2)}=-\frac{1}{\gamma \,
\theta_2(0)\theta_3(0)}\,\,
\frac{\theta_4(u_1)\theta_4(u_2)}{\theta_1(u_1)\theta_1(u_2)}\,\,
\frac{\theta_1(u_1-u_2)}{\theta_4(u_1-u_2)},
$$
where $u_i\equiv u(z_i)$. Using this identity, one can see that
the uniformization (\ref{E1}), (\ref{E2}) converts 
equations (\ref{D1a}), (\ref{D2a}) into a single
equation:
\beq\label{E4}
\left (z_1^{-1}-z_2^{-1}\right ) e^{\nabla (z_1) \nabla (z_2)F}
=\frac{\theta_1(u(z_1)\! -\! u(z_2))}{\theta_4(u(z_1)\! -\! u(z_2))}\,.
\eeq
This equation should be compared with (\ref{b3a}) which can be regarded as a limiting
case of (\ref{E4}) as $\tau \to +{\rm i}\infty$. 
Note that the limit $z_2\to \infty$ in (\ref{E4}) 
gives the definition of the 
function $u(z)$:
\beq\label{E6}
e^{\p_{t_0}\nabla (z)F}=z\,
\frac{\theta_1(u(z))}{\theta_4(u(z))}
\eeq
(which is equivalent to the first formula in (\ref{E1})).
Next, the $z\to \infty$ limit in (\ref{E6}) yields
\beq\label{E8}
R=e^{F_{00}}=\pi c_1 \, \theta_2(0)\theta_3(0),
\eeq
hence we conclude that
$c_1({\bf t})=\gamma ({\bf t})/\pi$.

Passing to logarithms and applying $\nabla (z_3)$ to the both sides of (\ref{E4}),
we obtain another useful form of this equation. 
It is convenient to introduce the function
\beq\label{E10}
S(u , \tau ):=\log \frac{\theta_1(u ,\tau )}{\theta_4(u ,\tau )}\,.
\eeq
It has the following quasiperiodicity properties:
$$S(u+1,\tau )=S(u,\tau )+i\pi, 
\quad S(u+\tau ,\tau )=S(u,\tau ).
$$
Below we write simply $S(u, \tau )=S(u)$ but it is important that
$S(u)$ depends on ${\bf t}$ not only through $u$ but also through $\tau$.
In terms of this function, the equation (\ref{E4}) means that
$\nabla (z_3) S (u(z_1)-u(z_2)) =\nabla (z_3)\nabla (z_2)\nabla (z_1)F$ is symmetric
under permutations of $z_1, z_2, z_3$:
\beq\label{E11}
\nabla (z_1) S (u(z_2)-u(z_3)) =
\nabla (z_2) S (u(z_1)-u(z_3)) =
\nabla (z_3) S (u(z_1)-u(z_2)) .
\eeq
In particular, as $z_3\to \infty$ we get
\beq\label{E12}
\nabla (z_1) S (u(z_2)) =\p_{t_0}S \Bigl (u(z_1)-u(z_2)\Bigr ).
\eeq
This is an elliptic extension of the first equation in (\ref{c2}). 
In the limit $z_2\to \infty$ equation (\ref{E12}) gives:
\beq\label{E12a}
\nabla (z)\log R= \p_{t_0} S\Bigl (u(z)\Bigr ).
\eeq

Equations (\ref{E11}) are in fact 
equivalent to the dDKP hierarchy in the form (\ref{E4}). Indeed,
it follows from (\ref{E12}) that $\nabla (z_1) S (u(z_2))=
\nabla (z_2) S (u(z_1))$. Therefore, there exists a function $f=f({\bf t})$ such that
$S(u(z))=\nabla (z)f$. Substituting this into (\ref{E12}) and integrating with respect to
$t_0$, we get
$$
\nabla (z_1)\nabla (z_2)F_1 +\log (z_1^{-1}-z_2^{-1})-S(u(z_1)-u(z_2))=c({\bf t}', z_1, z_2),
$$
where $\p_{t_0}F_1 = f$, $c({\bf t}', z_1, z_2)=
\displaystyle{\sum_{n,m\geq 0}c_{nm}({\bf t}')z_1^{-n}z_2^{-m}}$
is the integration constant and ${\bf t}'=\{t_1, t_2, \ldots \}$.
Applying $\nabla (z_3)$ to the both sides of this equation and using the symmetry
of $\nabla (z_3)S(u(z_1)-u(z_2))$ under the permutations of $z_1, z_2, z_3$
we conclude that $\nabla (z_3)c({\bf t}', z_1, z_2)$ is also
symmetric. Therefore, there exists a function $g$ such that
$c({\bf t}', z_1, z_2)=\nabla (z_1)\nabla (z_2)g$. Then $F=F_1-g$ satisfies (\ref{E4}).
Thus we may say that the pair $u(z, {\bf t})$, $\tau ({\bf t})$ satisfying (\ref{E11})
is a solution to the dDKP hierarchy.

In order to connect the elliptic formulation with the algebraic one, we note that
\beq\label{E13}
\log w(z)=-2S(u(z) ), \qquad
p(z)=c_1 S'(u(z)),
\eeq
where $S'(u)\equiv \p_u S(u)$. See \cite{akh-zab:14-1} for details.

\subsubsection{The reduction}

As in the previous cases, the one-variable reduction
means that solutions of the hierarchy 
$u(z, {\bf t})$ and $\tau ({\bf t})$ depend on the times 
through a single variable $\lambda = \lambda ({\bf t})$:
$u(z, {\bf t})=u(z, \lambda ({\bf t}))$, 
$\tau ({\bf t})=\tau (\lambda ({\bf t}))$.  
The function
of two variables, $u(z, \lambda )$, can not be arbitrary.
Our goal is to characterize the class of functions 
$u(z, \lambda )$, $\tau (\lambda )$ that are consistent with
the structure of the hierarchy and can be used for one-variable 
reductions.
Below we obtain the consistency condition for one-variable reductions of the 
dDKP hierarchy in the elliptic form.
The calculations are much more involved than in the cases of the 
dKP, dBKP and dToda hierarchies.

Applying the chain rule of differentiation to 
$
S\Bigl (u(z, {\bf t})| \tau ({\bf t})\Bigr )=
S \Bigl ( u(z, \lambda ({\bf t}))\, |\, \tau (\lambda ({\bf t}))\Bigr )
$,
we get:
$$
\nabla (z_1)S(u(z_2))=[\nabla (z_1)\lambda ]\,
\Bigl (\p_{\lambda }u(z_2)S'(u(z_2))+\p_{\lambda }\tau \dot S(u(z_2))\Bigr ),
$$
where the explicit dependence on $\tau$ is taken into account.
Here and below 
\beq\label{one0}
\dot S(u)=\p_{\tau} S(u, \tau ).
\eeq
Next, we have, using (\ref{E12a}):
$$
\nabla (z_1)\lambda = \frac{d\lambda}{d\log R}\, \nabla (z_1)\log R
=\frac{d\lambda}{d\log R}\, \p_{t_0}S(u(z_1))
$$
$$
=\,\, \frac{d\lambda}{d\log R}\, \p_{t_0}\lambda 
\Bigl ( \p_{\lambda }u(z_1)S'(u(z_1))+\p_{\lambda }\tau \dot S(u(z_1))\Bigr ).
$$
In a similar way, we get:
$$
\p_{t_0}S\Bigl (u(z_1)\! -\! u(z_2)\Bigr )=\p_{t_0}\lambda 
\Bigl [
\Bigl (\p_{\lambda}u(z_1)\! -\! \p_{\lambda}u(z_2)\Bigr )
S'\Bigl (u(z_1)\! -\! u(z_2)\Bigr ) +\p_{\lambda }\tau \dot S
\Bigl (u(z_1)\! -\! u(z_2)\Bigr ) \Bigr ].
$$
The formulas simplify a bit if we choose $\lambda =\tau$ which we set in what follows.
Assuming that $\p_{t_0}\tau \neq 0$ identically, we arrive at
the following relation:
\beq\label{one1}
\begin{array}{c}
\Bigl [\p_{\tau}u(z_1)S'(u(z_1))+\dot S(u(z_1))\Bigr ]\,
\Bigl [\p_{\tau}u(z_2)S'(u(z_2))+\dot S(u(z_2))\Bigr ]
\\ \\
=\,\, \displaystyle{
\frac{d\log R}{d\tau}\Bigl [
\Bigl (\p_{\tau}u(z_1)\! -\! \p_{\tau}u(z_2)\Bigr )S' 
\Bigl (u(z_1)\! -\! u(z_2)\Bigr )
+\dot S  \Bigl (u(z_1)\! -\! u(z_2)\Bigr )\Bigr ].}
\end{array}
\eeq

To proceed further, we need to know $\dot S(u)$. 
It is given by equation (\ref{n4}) in Appendix A.
The proof can be found in \cite{akh-zab:14-1}, for the proof of similar
formulae see \cite{Takasaki01,ZZ12}.
The key observation is that
the substitutions
\beq\label{one5}
\left \{
\begin{array}{l}
4\pi i \, \p_{\tau}u=\zeta_1(\xi -u)+\zeta_4(\xi -u )-
\zeta_1(\xi )-\zeta_4(\xi ),
\\ \\
4\pi i \, \p_{\tau}\log R =(S'(\xi ))^2,
\end{array} \right.
\eeq
where $\xi$ is an arbitrary parameter, convert equation 
(\ref{one1}) into identity. It is a highly nontrivial statement. 
Here we omit the details which
can be found in \cite{akh-zab:14-1}. 

Therefore, using the identity $\zeta_1(u,\tau )+\zeta_4(u, \tau )=
\zeta_1(u, \frac{\tau}{2} )$,
we conclude that the function $u(z, \tau )$ is compatible 
with the infinite hierarchy if it satisfies the differential equation
\beq\label{one6a}
4\pi i \, \p_{\tau}u(z)=-\zeta_1\Bigl (u(z)-\xi (\tau ), \frac{\tau}{2}\Bigr )-
\zeta_1\Bigl (\xi (\tau ), \frac{\tau}{2}\Bigr ),
\eeq
where $\xi (\tau)$ can be arbitrary real-valued function of $\tau$. 
This is essentially the Komatu-L\"owner equation
\eqref{komatu-loewner}. The variables in \S\ref{subsec:komatu-loewner}
correspond as
\[
    \tau = 2 \frac{\log q}{\pi i},\quad
    u(z,\tau) = \frac{\log g(z,q)}{2\pi i},\quad
    \xi(\tau) = \frac{\log \Lambda(q)}{2\pi i}.
\]
The driving function $\Lambda(q)=e^{2\pi i\xi(\tau)}$, which encodes the
shape of the slit in the L\"owner theory, specifies the reduction.

Coming back to an arbitrary reduction variable $\lambda =\lambda(\tau)$,
we obtain the equation
\beq\label{one6b}
\begin{array}{l}
4\pi i \, \p_{\lambda}u(z)=-\Bigl [\zeta_1\Bigl (u(z)-\xi (\lambda ), \frac{\tau}{2}\Bigr )+
\zeta_1\Bigl (\xi (\lambda ), \frac{\tau}{2}\Bigr )\Bigr ]\p_{\lambda}\tau .
\end{array}
\eeq
Equation (\ref{one6b}) written for the inverse function to the $u(z)$,
$z(u)$, reads
\beq\label{one6c}
\begin{array}{l}
4\pi i \, \p_{\lambda}z(u)=-\Bigl [\zeta_1\Bigl (u-\xi (\lambda ), \frac{\tau}{2}\Bigr )+
\zeta_1\Bigl (\xi (\lambda ), \frac{\tau}{2}\Bigr )\Bigr ]\p_u z(u) \, \p_{\lambda}\tau .
\end{array}
\eeq

One can also see that the second equation in (\ref{one5}), 
\beq\label{one7}
4\pi i \, \p_{\tau}\log R =(S'(\xi (\tau )))^2,
\eeq
emerges as the limiting case of (\ref{one6b}) when $z\to
\infty$.
Using some identities for the function $S$, it is possible to show that
for one-variable reductions the total $\tau$-derivative of $S(u(z))$
is given by
\beq\label{one8}
4\pi i \frac{dS(u(z))}{d\tau}=-S'(\xi (\tau ))\, 
S'\Bigl (u(z)-\xi (\tau )\Bigr ).
\eeq

Let us now pass to the system of reduced equations 
satisfied by $\lambda ({\bf t})$ and 
find their solution. 
Using (\ref{E12a}), we can write:
$$
\nabla (z) \tau = 
\frac{\p_{\tau}u(z) S'(u(z))+\dot S (u(z))}{d\log R/d\tau}\, \,
\p_{t_0}\tau = \frac{d S(u(z))/d\tau}{d\log R/d\tau}\, \,
\p_{t_0}\tau .
$$
Substituting (\ref{one7}), (\ref{one8}) and passing to $\lambda =\lambda (\tau)$, we get:
\beq\label{one9}
\nabla (z) \lambda = -\, \frac{S'\Bigl (u(z)-\xi (\lambda )\Bigr )}{S'(\xi (\lambda ))}
\, \, \p_{t_0}\lambda .
\eeq
This is a generating equation for a hierarchy of equations 
of the hydrodynamic type. To write them explicitly, we use the expansion
\beq\label{one10}
S'(u-u(z))=S'(u)+ \sum_{k\geq 1}\frac{z^{-k}}{k}\, B'_k(u)
\eeq
which defines the functions $B_k'(u)=B_k'(u, \tau )$. 
By analogy with the dispersionless KP case, 
one may think of them as elliptic analogues of the (derivatives of) Faber polynomials.
In terms of the functions $B_k'(u)$,
the equations of the reduced hierarchy are as follows:
\beq\label{one11}
\frac{\p \lambda}{\p t_k}= \phi_k (\xi (\lambda ), \tau )\, 
\frac{\p \lambda}{\p t_0}\,, \qquad
\phi_k (\xi (\lambda ), \tau ):=
\frac{B'_k(\xi (\lambda), \tau)}{S'(\xi (\lambda ), \tau)}\,, \quad
k\geq 1.
\eeq
The common solution to these equations can be written in the 
hodograph form:
\beq\label{one12}
t_0+\sum_{k=1}^{\infty} t_k \phi_k(\xi (\lambda ), \tau)=
\sfR(\lambda ),
\eeq
where $\sfR(\lambda )$ is an arbitrary function of $\lambda$. 

\section{Multivariable reductions}

\label{section:multi}

\subsection{Multivariable reductions of the dKP hierarchy}

\label{section:multidkp}

$N$-variable reduction of the dKP hierarchy
means that the function $p(z)$ depends on the times ${\bf t}$ 
through $N$ functions $\lambda_j=\lambda_j({\bf t})$, i.e., 
$p(z)=p(z, {\bf t})=p(z; \lambda_1({\bf t}), \ldots , \lambda_N({\bf t}))$. 
Below we prove that solutions of a system of L\"owner 
equations give solutions to the dKP hierarchy.

We consider the system of $N$ chordal L\"owner equations of the form 
(\ref{a9}) which 
characterize the dependence of $p(z)=p(z;\lambda_1, \ldots , \lambda_N)$ 
on the variables $\lambda_j$:
\beq\label{mrkp1}
\frac{\p p(z)}{\p \lambda_j}=-\, \frac{1}{p(z)-\xi_j}\, 
\frac{\p u}{\p \lambda_j}.
\eeq
The real-valued driving functions $\xi_j$ are functions of $\lambda_1, \ldots , \lambda_N$:
$\xi_j=\xi_j(\{\lambda_i\})$. 
The compatibility condition of the system (\ref{mrkp1}) is
$$
\frac{\p }{\p \lambda_j}\, \frac{\p p(z)}{\p \lambda_k}-
\frac{\p }{\p \lambda_k}\, \frac{\p p(z)}{\p \lambda_j}=0.
$$
A straightforward calculation (which uses the L\"owner
equations (\ref{mrkp1})) shows that it is equivalent to the following
Gibbons-Tsarev system:
\beq\label{mrkp2}
\left \{ \begin{array}{l}
\displaystyle{\frac{\p \xi_j}{\p \lambda_k}=\frac{1}{\xi_k-\xi_j}\,
\frac{\p u}{\p \lambda_k},}
\\ \\
\displaystyle{\frac{\p^2 u}{\p \lambda_j \p \lambda_k}=\frac{2}{(\xi_i-\xi_k)^2}\,
\frac{\p u}{\p \lambda_j}\, \frac{\p u}{\p \lambda_k}, \quad j\neq k}.
\end{array}\right.
\eeq

Now assume that each $\lambda_j$ is a function 
of the time variables. Then for 
$p(z, {\bf t})=p(z; \lambda_1({\bf t}), \ldots , \lambda_N({\bf t}))$
we can rewrite equation (\ref{a6}) in the form
$$
-\sum_{j=1}^N D(z_2)\lambda_j \cdot \p_{\lambda_j}p(z_1)=
\sum_{j=1}^N \p_{t_1}\lambda_j \cdot \p_{\lambda_j}\log \Bigl (p(z_1)-p(z_2)\Bigr ).
$$
Using the L\"owner equations (\ref{mrkp1}), we find:
$$
\p_{\lambda_j}\log \Bigl (p(z_1)-p(z_2)\Bigr )=
\frac{\p_{\lambda_j}u}{(p(z_1)-\xi_j)(p(z_2)-\xi_j)},
$$
and thus in order for (\ref{a6}) to be satisfied it is sufficient that the
following equation holds: 
\beq\label{mrkp3}
\sum_{j=1}^N D(z_2)\lambda_j \, \frac{\p_{\lambda_j}u}{p(z_1)-\xi_j}=
\sum_{j=1}^N \p_{t_1}\lambda_j \, \frac{\p_{\lambda_j}u}{(p(z_1)-\xi_j)(p(z_2)-\xi_j)}.
\eeq
It is clear that if we introduce the dependence of the $\lambda_j$'s
on ${\bf t}$ by means of the relations
\beq\label{mrkp4}
D(z)\lambda_j = \frac{\p_{t_1}\lambda_j}{p(z)-\xi_j},
\eeq
then equation (\ref{mrkp3}) is satisfied identically. 

Using definition of the Faber polynomials (\ref{a13}), one can see that
equation (\ref{mrkp4}) contains an infinite system of partial differential equations
of hydrodynamic type:
\beq\label{mrkp5}
\frac{\p \lambda_j}{\p t_k}=\phi_{j,k}(\{\lambda_i\})\frac{\p \lambda_j}{\p t_1},
\qquad \phi_{j,k}=B_k'(\xi_j), \quad k\geq 1, j=1, \ldots , N.
\eeq
Note that $\phi_{j,1}=B_1'(\xi_j)=1$, so that equations (\ref{mrkp5}) at 
$k=1$ become trivial identities. 
The generating function of $\phi_{j,k}(\{\lambda_i\})$ is
\beq\label{mrkp6}
\sum_{k\geq 1}\phi_{j,k}(\{\lambda_i\})\frac{z^{-k}}{k}
=\frac{1}{p(z)-\xi_j}
:=Q\Bigl (p(z, \{\lambda_i\}), \xi_j(\{\lambda_i\})\Bigr ).
\eeq

We will show that 
the system (\ref{mrkp5}) is consistent and can be solved by
Tsarev's generalized hodograph method \cite{tsa:90}.
It can be directly verified that 
the compatibility condition of the system (\ref{mrkp5}) is 
\beq\label{mrkp7a}
\frac{\p_{\lambda_j}\phi_{i,n}}{\phi_{j,n}-\phi_{i,n}}=
\frac{\p_{\lambda_j}\phi_{i,n'}}{\phi_{j,n'}-\phi_{i,n'}}\qquad \mbox{for all $i\neq j$, $n, n'$}.
\eeq
In other words the condition is
that 
\beq\label{mrkp7}
\Gamma_{ij}:=\frac{\p_{\lambda_j}\phi_{i,n}}{\phi_{j,n}-\phi_{i,n}}
\eeq
does not depend on $n$. It is easy to see that this is equivalent to the $z$-independence of the ratio
\beq\label{mrkp8}
\Gamma_{ij}=
\frac{\p_{\lambda_j}Q(p(z), \xi_i)}{Q(p(z), \xi_j)-Q(p(z), \xi_i)}=
\frac{\p_{\lambda_j}u}{(\xi_i-\xi_j)^2},
\eeq
where $Q$ is the generating function (\ref{mrkp6}). The 
proof of the second equality  
uses the L\"owner equation (\ref{mrkp1}) and the Gibbons-Tsarev equations (\ref{mrkp2}). 

Let $\sfR_i=\sfR_i(\{\lambda _j\})$
($i=1,\ldots,N$) satisfy the system of equations  
\beq\label{mrkp9}
\frac{\p \sfR_i}{\p \lambda_j}
=
\Gamma_{ij}(\sfR_j-\sfR_i), \qquad
i,j=1, \ldots , N, \quad i\neq j,
\eeq
where $\Gamma_{ij}$ is defined in (\ref{mrkp8}) (for $N=1$ this condition is void). We
claim that
the system (\ref{mrkp9}) is compatible in the sense of Tsarev \cite{tsa:90}.
To see this, we note that $\Gamma_{ij}$ (\ref{mrkp8})
can be expressed as
\beq\label{mrkp10}
\Gamma_{ij}=\frac{1}{2}\, \frac{\p }{\p \lambda_j}\log g_i, \quad
g_i=\frac{\p u}{\p \lambda_i}.
\eeq
It then follows that 
\beq\label{mrkp11}
\frac{\p \Gamma_{ij}}{\p \lambda_k}=\frac{\p \Gamma_{ik}}{\p \lambda_j}, \qquad i\neq j\neq k,
\eeq
which are Tsarev's compatibility conditions. This means that the system (\ref{mrkp5}) is semi-Hamiltonian.
The main geometric object associated with a semi-Hamiltonian system is a diagonal metric.
The quantities $g_i=g_{ii}$ are components of this metric while $\Gamma_{ij}=\Gamma_{ij}^{i}$ 
are the corresponding Christoffel symbols.
In fact the metric $g_i$ is of Egorov type, i.e., it holds
\beq\label{mrkp12}
\frac{\p g_i}{\p \lambda_k}=\frac{\p g_k}{\p \lambda_i}
\eeq
(this follows from (\ref{mrkp10})).

Assume that $\sfR_i$ satisfy the system (\ref{mrkp9}). 
Then the same argument as 
in the proof of Theorem 10 of Tsarev's paper \cite{tsa:90} shows that
if $\lambda_i({\bf t})$ is defined implicitly by the
hodograph relation
\beq\label{mrkp13}
t_1+\sum_{n\geq 2}\phi_{i,n}(\{\lambda_j\})t_n = 
\sfR_i(\{\lambda_j\}),
\eeq
then $\lambda_j({\bf t})$ satisfy (\ref{mrkp5}).

We have found sufficient conditions for $N$-variable 
diagonal reductions of the dKP hierarchy. The reduction is given by a system of $N$ chordal
L\"owner equations (\ref{mrkp1}) for a function $p(z, \lambda_1, \ldots , \lambda_N)$
supplemented by a diagonal system of hydrodynamic type (\ref{mrkp5}) 
for the variables 
$\lambda_j$, $j=1, \ldots , N$.

\subsection{Multivariable reductions of the dBKP hierarchy}

\label{section:multidbkp}

We consider the system of $N$ quadrant L\"owner equations of the form 
(\ref{bkp11}) which 
characterize the dependence of $p(z)=
p(z; \{\lambda_j\})$ on the variables $\lambda_j$:
\beq\label{mrbkp1}
\frac{\p p(z)}{\p \lambda_j}=-\, \frac{p(z)}{p^2(z)-\xi_j^2}\, 
\frac{\p u}{\p \lambda_j}.
\eeq
The compatibility conditions of this system are given by the Gibbons-Tsarev equations 
which have the form \cite{T13}
\beq\label{mrbkp2}
\left \{ \begin{array}{l}
\displaystyle{\frac{\p \xi_i^2}{\p \lambda_j}=\frac{2\xi_i^2}{\xi_j^2-\xi_i^2}\,
\frac{\p u}{\p \lambda_j},}
\\ \\
\displaystyle{\frac{\p^2 u}{\p \lambda_i \p \lambda_j}=
\frac{2(\xi_i^2+\xi_j^2)}{(\xi_i^2-\xi_j^2)^2}\,
\frac{\p u}{\p \lambda_i}\, \frac{\p u}{\p \lambda_j}, \quad i\neq j}.
\end{array}\right.
\eeq

Now assume that each
$\lambda_j$ is a function of ${\bf t}$:
$p(z;{\bf t}) = p(z;\{\lambda_j({\bf t})\})$.%
Then we can rewrite
equation (\ref{bkp7}) of the dBKP hierarchy for $p(z;{\bf t})$ in the form
$$
2\sum_{j=1}^N D^{\rm o}(z_2)\lambda_j \cdot \p_{\lambda_j}p(z_1)=
\sum_{j=1}^N \p_{t_1}\lambda_j \cdot \p_{\lambda_j}\log \frac{p(z_1)+p(z_2)}{p(z_1)-p(z_2)}.
$$
Using the quadrant L\"owner equations (\ref{mrbkp1}), we find:
$$
\p_{\lambda_j}\log \frac{p(z_1)+p(z_2)}{p(z_1)-p(z_2)}=
-\frac{2p(z_1)p(z_2)\p_{\lambda_j}u}{(p^2(z_1)-\xi_j^2)(p^2(z_2)-\xi_j^2)},
$$
and thus in order for (\ref{bkp7}) to be satisfied it is sufficient that
the following equation holds: 
\beq\label{mrbkp3}
\sum_{j=1}^N D^{\rm o}(z_2)\lambda_j \, \frac{p(z_1)\p_{\lambda_j}u}{p^2(z_1)-\xi_j^2}=
\sum_{j=1}^N \p_{t_1}\lambda_j \, \frac{p(z_1)p(z_2)\p_{\lambda_j}u}{(p^2(z_1)-
\xi_j^2)(p^2(z_2)-\xi_j^2)}.
\eeq
It is clear that if we introduce the dependence of the $\lambda_j$'s
on ${\bf t}$ by means of the relations
\beq\label{mrbkp4}
D^{\rm o}(z)\lambda_j = \frac{p(z)\, \p_{t_1}\lambda_j}{p^2(z)-\xi_j^2},
\eeq
then equation (\ref{mrbkp3}) is satisfied identically. Since 
$$
\frac{p(z)}{p^2(z)-\xi_j^2}=\sum_{k\geq 1, {\rm odd}}\frac{z^{-k}}{k}\, B_k'(\xi_j)
$$
(see (\ref{bkp9a})), we see that
equation (\ref{mrbkp4}) contains an infinite system of partial differential equations
of hydrodynamic type which have the same form as (\ref{mrkp5}) with the only difference 
that $k$ is odd:
\beq\label{mrbkp5}
\frac{\p \lambda_j}{\p t_k}=\phi_{j,k}(\{\lambda_i\})\frac{\p \lambda_j}{\p t_1},
\quad \phi_{j,k}=B_k'(\xi_j), \quad k=1, 3, 5, \ldots \, , \quad j=1, \ldots , N.
\eeq
The generating function of $\phi_{j,k}(\{\lambda_i\})$ is
\beq\label{mrbkp6}
\sum_{k\geq 1, \, {\rm odd}}\phi_{j,k}(\{\lambda_i\})\frac{z^{-k}}{k}
=\frac{p(z)}{p^2(z)-\xi_j^2}
:=Q\Bigl (p(z, \{\lambda_i\}), \xi_j(\{\lambda_i\})\Bigr ).
\eeq

The compatibility condition of
the system (\ref{mrbkp5}) has the same form (\ref{mrkp7a}) as before, which means
that $\Gamma_{ij}$ given by (\ref{mrkp7}) does not depend on $n$. 
Again, this is equivalent to the $z$-independence of 
\beq\label{mrbkp8}
\Gamma_{ij}=
\frac{\p_{\lambda_j}Q(p(z), \xi_i)}{Q(p(z), \xi_j)-Q(p(z), \xi_i)}=
\frac{\xi_i^2+\xi_j^2}{(\xi_i^2-\xi_j^2)^2} \, \frac{\p u}{\p \lambda_j},
\eeq
where $Q$ is the generating function (\ref{mrbkp6}). 

Let $R_i=R_i(\{\lambda _j\})$ ($i=1, \ldots , N$) satisfy the system
(\ref{mrkp9}),
where $\Gamma_{ij}$ is defined in (\ref{mrbkp8}). 
Tsarev's compatibility conditions (\ref{mrkp11}) follow from (\ref{mrkp10}) which 
remains of the
same form and the same argument is used to show
that if $\lambda_i({\bf t})$ is defined implicitly by the
hodograph relation
\beq\label{mrbkp13}
t_1+\sum_{n\geq 3, \, {\rm odd}}\phi_{i,n}(\{\lambda_j\})t_n = R_i(\{\lambda_j\}),
\eeq
then $\lambda_j({\bf t})$ satisfy (\ref{mrbkp5}).

\subsection{Multivariable reductions of the dToda hierarchy}

\label{section:multidtoda}

We consider the system of $N$ radial L\"owner equations of the form 
(\ref{b15}) which 
characterize the dependence of $w(z)=w(z; \{\lambda_j\})$ on the variables $\lambda_j$:
\beq\label{mrt1}
\frac{\p \log w(z)}{\p \lambda_j}=-\, \frac{w(z)+\eta_j}{w(z)-\eta_j}\, 
\frac{\p \log r}{\p \lambda_j}, \qquad \eta_j=e^{i\xi_j}.
\eeq
The driving functions $\xi_j=\xi_j(\{\lambda_i\})$ are real-valued. 
The compatibility conditions 
$$\displaystyle{\frac{\p }{\p \lambda_k}\frac{\p \log w(z)}{\p \lambda_j}=
\frac{\p }{\p \lambda_j}\frac{\p \log w(z)}{\p \lambda_k}}$$ 
of this system are given by the Gibbons-Tsarev equations:
\beq\label{mrt2}
\left \{ \begin{array}{l}
\displaystyle{\frac{\p \eta_j}{\p \lambda_k}=\eta_j \, \frac{\eta_k+\eta_j}{\eta_k-\eta_j}
\, \frac{\p \log r}{\p \lambda_k}},
\\ \\
\displaystyle{\frac{\p^2 \log r}{\p \lambda_j \p \lambda_k}=
\frac{4\eta_j\eta_k}{(\eta_j-\eta_k)^2}\, \frac{\p \log r}{\p \lambda_j}\,
\frac{\p \log r}{\p \lambda_k}}
\end{array}\right.
\eeq
or
\beq\label{mrt2a}
\left \{ \begin{array}{l}
\displaystyle{\frac{\p \xi_j}{\p \lambda_k}=\cot \Bigl (\frac{1}{2}\,
(\xi_j-\xi_k)\Bigr )\,
\frac{\p \log r}{\p \lambda_k}},
\\ \\
\displaystyle{\frac{\p^2 \log r}{\p \lambda_j \p \lambda_k}=-\,
\frac{1}{\sin^2\Bigl (\frac{1}{2}(\xi_j-\xi_k)\Bigr )}\,
\frac{\p \log r}{\p \lambda_j}\, \frac{\p \log r}{\p \lambda_k}, \quad j\neq k}.
\end{array}\right.
\eeq

Now assume that each $\lambda_j$
is a function of the times:
$w(z;{\bf t})=w(z;\{\lambda_j({\bf t})\})$.
Then we can rewrite equations (\ref{c2}) of the dToda
hierarchy in the form
$$
-\sum_{j=1}^N D(z_1)\lambda_j \cdot \p_{\lambda_j}\log w(z_2)=
\! \sum_{j=1}^N\p_{t_0}\lambda_j \left [
\p_{\lambda_j}\log \Bigl (w(z_1)\! -\! w(z_2)\Bigr ) -\frac{1}{2}\, \p_{\lambda_j}
\log \frac{w(z_1)}{r}\right ],
$$
$$
\sum_{j=1}^N \bar D(\bar z_1)\lambda_j \cdot \p_{\lambda_j}\log w(z_2)=
\! \sum_{j=1}^N\p_{t_0}\lambda_j \left [
\p_{\lambda_j}\log \Bigl (\bar w(\bar z_1)\! -\! w^{-1}(z_2)\Bigr ) -
\frac{1}{2}\, \p_{\lambda_j}
\log \frac{\bar w(\bar z_1)}{r}\right ],
$$
Using the radial L\"owner equations (\ref{mrt1}), we can rewrite these relations as
\beq\label{mrt2b}
\sum_{j=1}^N D(z_1)\lambda_j \frac{w(z_2)+\eta_j}{w(z_2)-\eta_j}\, \p_{\lambda_j}
\log r =\sum_{j=1}^N\p_{t_0}\lambda_j\frac{\eta_j (w(z_2)+\eta_j)}{(w(z_1)-\eta_j)
(w(z_2)-\eta_j)}\, \p_{\lambda_j}\log r,
\eeq
\beq\label{mrt2c}
-\sum_{j=1}^N \bar D(\bar z_1)\lambda_j 
\frac{w(z_2)+\eta_j}{w(z_2)-\eta_j}\, \p_{\lambda_j} \log r =\sum_{j=1}^N\p_{t_0}\lambda_j
\frac{w(z_2)+\eta_j}{(1-\eta_j \bar w(\bar z_1))
(w(z_2)-\eta_j)}\, \p_{\lambda_j}\log r.
\eeq
It is clear that if we introduce the dependence of the $\lambda_j$'s
on ${\bf t}, \bar {\bf t}$ by means of the relations
\beq\label{mrt3}
\begin{array}{l}
\displaystyle{
D(z)\lambda_j=\frac{\eta_j}{w(z)-\eta_j}\, \p_{t_0}\lambda_j},
\\ \\
\displaystyle{\bar D(\bar z)\lambda_j=\frac{1}{\eta_j\bar w(\bar z)-1}\, \p_{t_0}\lambda_j},
\end{array}
\eeq
then equations (\ref{mrt2b}), (\ref{mrt2c}) are satisfied identically. Note that the two
equations (\ref{mrt3}) are complex conjugate to each other (recall that $\bar \eta_j =
\eta_j^{-1}$). 

As it follows from (\ref{b17a}), 
equation (\ref{mrt3}) contains an infinite system of partial differential equations
of hydrodynamic type:
\beq\label{mrt4}
\frac{\p \lambda_j}{\p t_k}=\phi_{j,k}(\{\lambda_i\})\frac{\p \lambda_j}{\p t_0},
\quad \phi_{j,k}=\eta_jA_k'(\eta_j), \quad k\geq 1\, , \quad j=1, \ldots , N.
\eeq
The generating function of $\phi_{j,k}(\{\lambda_i\})$ is
\beq\label{mrt5}
\sum_{k\geq 1}\phi_{j,k}(\{\lambda_i\})\frac{z^{-k}}{k}
=\frac{\eta_j}{w(z)-\eta_j}
:=Q\Bigl (w(z, \{\lambda_i\}), \eta_j(\{\lambda_i\})\Bigr ).
\eeq
The compatibility condition of
the system (\ref{mrt4}) has the same form as before. As a straightforward calculation 
shows, 
\beq\label{mrt6}
\Gamma_{ij}=
\frac{\p_{\lambda_j}Q(w(z), \eta_i)}{Q(w(z), \eta_j)-Q(w(z), \eta_i)}=
\frac{2\eta_i\eta_j}{(\eta_i-\eta_j)^2}\, \p_{\lambda_j}\! \log r,
\eeq
does not depend on $z$. 

Let $R_i=R_i(\{\lambda _j\})$ ($i=1, \ldots , N$) satisfy the system (\ref{mrkp9}),
where $\Gamma_{ij}$ is defined in (\ref{mrt6}). 
Tsarev's compatibility conditions (\ref{mrkp11}) follow from the fact that
\beq\label{mrt7}
\Gamma_{ij}=\frac{1}{2}\, \frac{\p }{\p \lambda_j}\log g_i, \quad
g_i=\frac{\p \log r}{\p \lambda_i}
\eeq
which is easily checked using the Gibbons-Tsarev equation (\ref{mrt2}).
Finally, the same argument as before is used to show
that if $\lambda_i({\bf t})$ is defined implicitly by the
hodograph relation
\beq\label{mrt8}
t_0+2{\rm Re}\sum_{n\geq 1, }\phi_{i,n}(\{\lambda_j\})t_n = R_i(\{\lambda_j\}),
\eeq
then $\lambda_j({\bf t})$ satisfy (\ref{mrt4}).

\subsection{Multivariable reductions of the dDKP hierarchy in the elliptic form}

\label{section:multiddkp}

In this section we study diagonal $N$-variable reductions of the dDKP hierarchy
in the elliptic form when the function 
$u=u(z)$ depends on the times ${\bf t}$ through $N$ real variables $\lambda_j$:
$u(z; {\bf t})=u(z; \{\lambda_j({\bf t})\})$.
Following \cite{ATZ17}, we prove that solutions of a system of elliptic L\"owner 
equations give solutions to the dDKP hierarchy.

\subsubsection{The Gibbons-Tsarev equations and a system of hydrodynamic type}

The starting point is the system of $N$ elliptic L\"owner equations which 
characterize the dependence of $u(z)$ on the variables $\lambda_j$:
\beq
\label{gt2}
\begin{array}{lll}
\displaystyle{\frac{\p u}{\p \lambda_j}}&=&\displaystyle{-\frac{1}{4\pi {\rm i}}\Bigl (
\zeta_1 (u-\xi_j, \tau)+\zeta_4 (u-\xi_j,\tau )+\zeta_1(\xi_j,\tau )+\zeta_4(\xi_j, \tau )\Bigr )
\frac{\p \tau}{\p \lambda_j}}
\\ &&\\
&=&\displaystyle{-\frac{1}{4\pi {\rm i}}\Bigl (
\zeta_1 (u-\xi_j, \tau ')+\zeta_1(\xi_j,\tau ' )\Bigr )
\frac{\p \tau}{\p \lambda_j}}.
\end{array}
\eeq
Here and below we abbreviate $\tau ' =\frac{\tau}{2}$.
In (\ref{gt2}), $u=u(z, \{\lambda_i\})$ is a function of $z$ and real variables
$\{\lambda_i\}=\{\lambda_1, \ldots , \lambda_N\}$ which are functions of the times. 
The real-valued driving functions are
$\xi_j=\xi_j(\{\lambda_i\})$. 

The Gibbons-Tsarev system is the compatibility condition for the system 
of elliptic L\"owner equations (\ref{gt2}):
$$
\frac{\p }{\p \lambda_j}\, \frac{\p u}{\p \lambda_k}-
\frac{\p }{\p \lambda_k}\, \frac{\p u}{\p \lambda_j}=0.
$$
As is shown in \cite{ATZ17}, this compatibility condition is equivalent 
to the elliptic analogue of the Gibbons-Tsarev system:
\beq\label{GT1}
\frac{\p \xi_k}{\p \lambda_j}=\frac{1}{4\pi {\rm i}}\, \Bigl (
\zeta_1(\xi_j-\xi_k, \tau ')-\zeta_1(\xi_j, \tau ')\Bigr )\, \frac{\p \tau}{\p \lambda_j},
\eeq
\beq\label{GT2}
\frac{\p^2 \tau}{\p \lambda_k \p \lambda_j}=\frac{1}{2\pi {\rm i}}\,
\wp _1(\xi_k-\xi_j, \tau ')\frac{\p \tau}{\p \lambda_k}\frac{\p \tau}{\p \lambda_j}
\eeq
for all $j=1, \ldots , N$, $j\neq k$. 
The system of 
equations (\ref{GT1}), (\ref{GT2}) is the elliptic analogue of the famous Gibbons-Tsarev system
\cite{GT1,GT2}. They already appeared in the literature 
\cite{ode-sok:09,ode-sok:10,ode-sok:09-1}.

Let us calculate $\p_{\lambda_j}S(u(z_1)-u(z_2))$:
$$
\p_{\lambda_j}S(u_1-u_2)=
S'(u_1-u_2)\Bigl (\frac{\p u_1}{\p \lambda_j}-\frac{\p u_2}{\p \lambda_j}\Bigr )
+\dot S(u_1-u_2)\frac{\p \tau}{\p \lambda_j}\,,
$$
where we abbreviate $u_i \equiv u(z_i)$ and $\dot S(u)=\p_{\tau}S(u, \tau )$.
Plugging here the elliptic L\"owner equations (\ref{gt2}) and the formula (\ref{n4})
for $\dot S(u)$, 
we have:
\beq\label{f201}
\begin{array}{c}
\p_{\lambda_j}S(u_1\! -\! u_2)=
\displaystyle{\frac{1}{4\pi {\rm i}}S'(u_1-u_2) \Bigl [
-\zeta_1 (u_1-\xi_j)-\zeta_4 (u_1-\xi_j)+\zeta_1 (u_2-\xi_j)+\zeta_4 (u_2-\xi_j)}
\\  \\
 \displaystyle{ +\, 2\zeta_2(u_1-u_2)+\frac{\pi^2 \theta_4^4(0, \tau)}{S'(u_1-u_2)}
\Bigr ] \frac{\p \tau}{\p \lambda_j}
=\frac{1}{4\pi {\rm i}} S'(u_1 -\xi_j)S'(u_2-\xi_j)
\frac{\p \tau}{\p \lambda_j}.
}
\end{array}
\eeq
Here we have used identity (A15) from \cite{akh-zab:14-1}.
In particular, tending $z_2\to \infty$, we get
\beq\label{f202}
\p_{\lambda_j}S(u(z))=-\frac{1}{4\pi {\rm i}} S'(\xi_j)S'(u(z)-\xi_j)
\frac{\p \tau}{\p \lambda_j}.
\eeq

Let us construct
a solution $u(z)$, $\tau$ to the dDKP hierarchy which depends on times through the 
$\lambda_i$'s: $u(z, {\bf t})=u(z, \{\lambda_i ({\bf t})\})$, 
$\tau ({\bf t})=\tau (\{\lambda_i({\bf t})\})$.
It is an $N$-variable reduction of the hierarchy.
If the reduction condition is imposed,
main equation (\ref{E12}) reads
$$
\sum_{j=1}^N\nabla (z_1)\lambda_j \cdot
\p_{\lambda_j}S(u(z_2))=\sum_{j=1}^N \p_{t_0}\lambda_j \cdot
\p_{\lambda_j}S(u(z_1)-u(z_2)).
$$
Plugging here equations (\ref{f201}), (\ref{f202}), we obtain:
\beq\label{f203}
-\sum_{j=1}^N\nabla (z_1)\lambda_j \cdot S'(\xi_j)S'(u(z_2)\! -\! \xi_j)
\frac{\p \tau}{\p \lambda_j}
=\sum_{j=1}^N\p_{t_0}\lambda_j \cdot S'(u(z_1)\! -\! \xi_j)S'(u(z_2)\! -\! \xi_j)
\frac{\p \tau}{\p \lambda_j}.
\eeq
It is clear from (\ref{f203}) 
that if we introduce the dependence of the $\lambda_j$'s on ${\bf t}$ 
by means of the relation
\beq\label{f3}
\nabla (z)\lambda_j=-\frac{S'(u(z)-\xi_j)}{S'(\xi_j)}\, \frac{\p \lambda_j}{\p t_0},
\eeq
equation (\ref{f203}) is satisfied identically. 
It is easy to see that equations (\ref{E11})
are also satisfied. 
Indeed, 
using the reduction condition, we rewrite equation (\ref{E11}) as
$$
\sum_{j=1}^N\nabla (z_3)\lambda_j \cdot
\p_{\lambda_j}S(u(z_1)-u(z_2))\,\,\,\,
\mbox{is symmetric under permutations of }z_1,z_2,z_3.
$$
Plugging here (\ref{f201}), we have that
$\displaystyle{
\sum_{j=1}^N\nabla (z_3)\lambda_j \cdot S'(u(z_1)\! -\! \xi_j)
S'(u(z_2)\! -\! \xi_j)}$
is symmetric under permutations of $z_1,z_2,z_3$ if
$\nabla (z) \lambda_j$ is given by (\ref{f3}).

Equation (\ref{f3}) contains an infinite system of partial differential equations 
of hydrodynamic type. To write them out explicitly, we use 
the (derivatives of) elliptic Faber functions 
$B_k' (u)$ introduced via the expansion (\ref{one10}). 
Then the system (\ref{f3}) reads
\beq\label{f5}
\frac{\p \lambda _j}{\p t_k}=\phi_{j,k}(\{\lambda_i\})\, \frac{\p \lambda _j}{\p t_0}\,, \qquad
\phi_{j,k}=\frac{B '_k(\xi_j)}{S'(\xi_j)},
\eeq
which is an infinite diagonal system of partial differential equations of hydrodynamic type.
The $\lambda_j$'s play the role of the Riemann invariants. 
The generating function of $\phi_{j,k}(\{\lambda_i\})$ is
\beq\label{f6}
1+\sum_{k\geq 1}\phi_{j,k}(\{\lambda_i\})\frac{z^{-k}}{k}
:=Q\Bigl (u(z, \{\lambda_i\}), \xi_j(\{\lambda_i\}), \tau (\{\lambda_i\})\Bigr ),
\eeq
$$
Q(u, \xi , \tau )=\frac{S'(\xi -u , \tau )}{S'(\xi , \tau )}.
$$
It is convenient to put $\phi_{i,0}=1$.

\subsubsection{Generalized hodograph method}

We have reduced the dDKP hierarchy to
the system of elliptic L\"owner equations and the auxiliary equations
of hydrodynamic type
\begin{equation}\label{g0}
    \frac{\p \lambda_i ({\bf t})}{\p t_n}
    =
    \phi_{i,n}(\{\lambda_j({\bf t})\}) \frac{\p\lambda_i({\bf t})}{\p t_0},
\end{equation}
where $\phi_{i,n}$ are given by (\ref{f5}).
This system is consistent and can be solved by
Tsarev's generalized hodograph method \cite{tsa:90} as were the previous cases 
(see sections \ref{section:multidkp}, \ref{section:multidbkp}, \ref{section:multidtoda}).

It can be directly verified that 
the compatibility condition of the system (\ref{g0}) is 
$$
\frac{\p_{\lambda_j}\phi_{i,n}}{\phi_{j,n}-\phi_{i,n}}=
\frac{\p_{\lambda_j}\phi_{i,n'}}{\phi_{j,n'}-\phi_{i,n'}}\qquad \mbox{for all $i\neq j$, $n, n'$}.
$$
In other words the condition is
that 
\beq\label{g1}
\Gamma_{ij}:=\frac{\p_{\lambda_j}\phi_{i,n}}{\phi_{j,n}-\phi_{i,n}}
\eeq
does not depend on $n$, or, equivalently that
\beq\label{g2}
\Gamma_{ij}=
\frac{\p_{\lambda_j}Q(u(z), \xi_i, \tau )}{Q(u(z), \xi_j, \tau )-Q(u(z), \xi_i, \tau )}
\eeq
does not depend on $z$,
where $Q$ is the generating function (\ref{f6}).
The independence of (\ref{g2}) of $z$ is proved in 
\cite{ATZ17} by a direct cumbersome calculation  
and the coefficients $\Gamma_{ij}$ are found:
\beq\label{g3}
\Gamma_{ij}=-\, \frac{1}{4\pi {\rm i}}\, \frac{S'(\xi_j)}{S'(\xi_i)}
\, S''(\xi_i-\xi_j)\, \frac{\p \tau}{\p \lambda_j}.
\eeq

Let $\sfR_i=\sfR_i(\{\lambda _j\})$ ($i=1, \ldots , N$) 
satisfy the system of equations 
\beq\label{g4}
\frac{\p \sfR_i}{\p \lambda_j}=\Gamma_{ij}(\sfR_j-\sfR_i), \qquad
i,j=1, \ldots , N, \quad i\neq j,
\eeq
where $\Gamma_{ij}$ is defined in (\ref{g3}) (for $N=1$ this condition is void). We
claim that
the system (\ref{g4}) is compatible in the sense of Tsarev \cite{tsa:90}.
To see this, we note that $\Gamma_{ij}$ (\ref{g3})
can be expressed as logarithmic derivative of a function:
\beq\label{c201}
\Gamma_{ij}=\frac{1}{2}\, \frac{\p }{\p \lambda_j}\log g_i, \quad
g_i=\frac{1}{4\pi {\rm i}}\, (S'(\xi_i))^2 \, \frac{\p \tau}{\p \lambda_i}.
\eeq
This is proved in \cite{ATZ17}. It then follows that 
\beq\label{c2a1}
\frac{\p \Gamma_{ij}}{\p \lambda_k}=\frac{\p \Gamma_{ik}}{\p \lambda_j}, \qquad i\neq j\neq k,
\eeq
which are Tsarev's compatibility conditions. This means that the system (\ref{g0}) is semi-Hamiltonian.
The main geometric object associated with a semi-Hamiltonian system is a diagonal metric.
The quantities $g_i=g_{ii}$ are components of this metric while $\Gamma_{ij}=\Gamma_{ij}^{i}$ 
are the corresponding Christoffel symbols.
In fact the metric $g_i$ is of Egorov type, i.e., it holds
\beq\label{q1}
\frac{\p g_i}{\p \lambda_k}=\frac{\p g_k}{\p \lambda_i}.
\eeq
The proof is very simple:
$$
\frac{\p g_k}{\p \lambda_i}=g_k\frac{\p \log g_k}{\p \lambda_i}
=2g_k\Gamma_{ki}=
\frac{1}{8\pi^2}\, S'(\xi_i)S'(\xi_k)S''(\xi_i-\xi_k)\frac{\p \tau}{\p \lambda_i}\,
\frac{\p \tau}{\p \lambda_k}
$$
from which we see that the right hand side
is explicitly symmetric under the permutation of $i$ and $k$.
(Here we have used (\ref{g3}) and (\ref{c201}).)

Assume that $\sfR_i$ satisfy the system (\ref{g4}). 
We claim that if $\lambda_i({\bf t})$ is defined implicitly by the
hodograph relation
\beq\label{g5}
t_0+\sum_{n\geq 1}\phi_{i,n}(\{\lambda_j\})t_n 
= \sfR_i(\{\lambda_j\}),
\eeq
then $\lambda_j({\bf t})$ satisfy (\ref{g0}).
The proof is almost 
the same as that of Theorem 10 of Tsarev's paper \cite{tsa:90}. 

To conclude, we have found sufficient conditions for $N$-variable 
diagonal reductions of the dDKP hierarchy
in the elliptic 
parametrization. The reduction is given by a system of $N$ elliptic
L\"owner equations (\ref{gt2}) for a function $u(z, \lambda_1, \ldots , \lambda_N)$
supplemented by a diagonal system of hydrodynamic type (\ref{f5}) 
for the variables 
$\lambda_j$, $j=1, \ldots , N$. We have derived compatibility conditions for the 
elliptic
L\"owner equations which are elliptic analogues of the Gibbons-Tsarev equations and
have
proved solvability of the hydrodynamic type system by means of the 
generalized hodograph method.

\section*{Acknowledgments}

\addcontentsline{toc}{section}{Acknowledgements}

We thank I. Hotta for valuable correspondence and K. Basarov for help in preparing 
figures.  The work of A.Z. was supported in part by 
RFBR grant 18-01-00461. The work of T.T. has been funded within the framework of the HSE 
University Basic Research Program and the Russian Academic Excellence Project '5-100'.

\subsection*{Appendix A: necessary functions and identities}
\def\theequation{A\arabic{equation}}
\setcounter{equation}{0}

\addcontentsline{toc}{section}{Appendix A: necessary functions and identities}

The Jacobi's theta-functions $\theta_a (u)=
\theta_a (u,\tau )$, $a=1,2,3,4$, are defined by the formulas
\beq\label{Bp1}
\begin{array}{l}
\theta _1(u)=-\displaystyle{\sum _{k\in \z}}
\exp \left (
\pi {\rm i} \tau (k+\frac{1}{2})^2 +2\pi {\rm i}
(u+\frac{1}{2})(k+\frac{1}{2})\right ),
\\
\theta _2(u)=\displaystyle{\sum _{k\in \z}}
\exp \left (
\pi {\rm i} \tau (k+\frac{1}{2})^2 +2\pi {\rm i}
u(k+\frac{1}{2})\right ),
\\
\theta _3(u)=\displaystyle{\sum _{k\in \z}}
\exp \left (
\pi {\rm i} \tau k^2 +2\pi {\rm i} u k \right ),
\\
\theta _4(u)=\displaystyle{\sum _{k\in \z}}
\exp \left (
\pi {\rm i} \tau k^2 +2\pi {\rm i}
(u+\frac{1}{2})k\right ),
\end{array}
\eeq where $\tau$ is a complex parameter (the modular parameter) 
such that ${\rm Im}\, \tau >0$. The function 
$\theta_1(u)$ is odd, the other three functions are even.
The infinite product representation for the $\theta_1(u)$ reads: 
\beq
\label{infprod} \theta_1(u)={\rm i}\,\mbox{exp}\, 
\Bigl ( \frac{{\rm i}\pi \tau}{4}-{\rm i}\pi u\Bigr ) \prod_{k=1}^{\infty} 
\Bigl (
1-e^{2\pi {\rm i} k\tau }\Bigr ) \Bigl ( 1-e^{2\pi {\rm i} ((k-1)\tau +u)}\Bigr ) 
\Bigl ( 1-e^{2\pi {\rm i} (k\tau -u)}\Bigr ). 
\eeq 
We also mention the identity
\beq\label{theta1prime}
\theta_1'(0)=\pi \theta_2(0) \theta_3(0) \theta_4(0).
\eeq
Many useful identities for the theta functions can be found in \cite{KZ1}.

All formulas for derivatives of elliptic functions with respect 
to the modular parameter follow from the ``heat equation'' satisfied
by the theta-functions:
\beq\label{heat}
4\pi {\rm i} \, \p_{\tau}\theta_a(u, \tau)=\p_u^2 \theta_a(u, \tau).
\eeq

In the main text we use the functions
\begin{equation}
\zeta_a (x, \tau)=\frac{\p }{\p x}\log \theta_a(x, \tau), 
\qquad
\wp_a(x, \tau )=-\frac{\p }{\p x}\zeta_a (x, \tau), \qquad a=1,2,3,4.
\label{zeta-wp} 
\end{equation}
Obviously, $\zeta_a$ are odd functions. In particular, 
$\zeta_1(x, \tau)=\frac{1}{x}+O(x)$ as $x\to 0$ and $\zeta_a(0, \tau)=0$ for $a=2,3,4$.
The functions $\zeta_1(x)=\zeta_1(x,\tau)$ and $\wp_1(x)=\wp(x, \tau)$
are up to some simple correction terms the Weierstrass $\zeta$- and $\wp$-functions
respectively. 

From the infinite product representation \eqref{infprod} follows the
explicit form of $\zeta_1(u)$:
\begin{equation}
    \zeta_1(u)
    =
    -\pi i\, \calK_{e^{\pi i \tau}}(e^{2\pi i u}),
    \quad
    \mathcal{K}_q(z)
    :=
    \frac{1+z}{1-z}
    +
    2 \sum_{k=1}^\infty \frac{q^{2k}z}{1-q^{2k}z}
    +
    2 \sum_{k=1}^\infty \frac{q^{2k} }{q^{2k}-z}.
\label{zeta1:expansion}
\end{equation}
The function $\calK_q(z)$ is called {\em Villat's function} ((2.2) of
\cite{fuk-kan:14}; cf.\ also \cite{Goluzin}). {\em Villat's kernel},
\begin{equation}
    \calK_q(z,\zeta):=\calK_q(z/\zeta)
\label{villat-kernel}
\end{equation}
plays an important role in the
L\"owner theory of doubly connected domains (\cite{Komatu},
\cite{fuk-kan:14}, \cite{Goluzin}).  

Let us introduce the function
\beq\label{n1}
S(x)=\log \frac{\theta_1(x, \tau)}{\theta_4(x,\tau )}.
\eeq
We denote $\p_x S(x)=S'(x)$, $\p^2_x S(x)=S''(x)$,
$\p_{\tau}S(x)=\dot S(x)$. One can prove the following formulae (here and below 
$\tau' \equiv \frac{\tau}{2}$):
\beq\label{n2}
S'(x)=\pi \theta_4^2(0, \tau )\, 
\frac{\theta_2(x, \tau)\theta_3(x,\tau )}{\theta_1(x, \tau)\theta_4(x,\tau )}
=\pi \theta_3(0,\tau')\theta_4(0,\tau')\, \frac{\theta_2(x, \tau')}{\theta_1(x, \tau')},
\eeq
\beq\label{n3}
\begin{array}{lll}
S''(x)&=&\displaystyle{-\pi^2\theta_2^2(0, \tau)\theta_3^2(0, \tau)\theta_4^3(0, \tau)\,
\frac{\theta_4(2x, \tau )}{\theta_1^2(x, \tau)\theta_4^2(x, \tau)}}
\\ &&\\
&=&\displaystyle{-\pi^2 \theta_3(0,\tau')\theta_4(0,\tau')\theta_2^2(0,\tau')\,
\frac{\theta_3(x, \tau')\theta_4(x, \tau')}{\theta_1^2(x,\tau')},}
\end{array}
\eeq
\beq\label{n4}
2\pi {\rm i}\dot S(x)=S'(x)\zeta_2 (x, \tau)+\frac{\pi^2}{2}\, \theta_4^4(0, \tau),
\eeq
\beq\label{n5}
2\pi {\rm i}\dot S'(x)=S''(x)\zeta_2(x, \tau)-S'(x)\wp_2(x, \tau ).
\eeq
It is clear from (\ref{n2}), (\ref{n3}) that
$
S'(x+1)=S'(x)$, $S'(x+\tau ')=-S'(x)$,
$S''(x+1)=S''(x)$, $S''(x+\tau ')=-S''(x)
$.
Note that
\beq\label{n6a}
\begin{array}{c}
S'(x)S'(x+\frac{1}{2})=-\pi^2\theta_4^4(0, \tau).
\end{array}
\eeq
Clearly, $S'(x)$ is an odd function. As $x\to 0$, we have:
$$
S'(x)=\frac{1}{x}+O(x), \qquad S''(x)=-\frac{1}{x^2}+O(1).
$$

\end{document}